\documentclass{jpp}
\usepackage{graphicx}
\usepackage{epstopdf, epsfig}
\usepackage{hyperref}
\usepackage{amsmath,amsfonts,amssymb}
\usepackage{subfigure}
\usepackage{relsize}
\usepackage{fontawesome}

\DeclareMathOperator{\sech}{sech}

\graphicspath{{sources/figs}}

\usepackage{color,soul}
\usepackage{hyperref}
\hypersetup{
  colorlinks   = true, %
  urlcolor     = blue, %
  linkcolor    = blue, %
  citecolor   = blue %
}

\newcommand{\Eq}[1]{Eq.~(\ref{#1})}
\newcommand{\Eqs}[2]{Eqs.~(\ref{#1}) and~(\ref{#2})}
\newcommand{\Fig}[1]{Fig.~\ref{#1}}

\shorttitle{Tearing modes in 3D}
\shortauthor{Vinay Kumar and Pallavi Bhat}

\title{Three-dimensional tearing instability of flux-tube-like magnetic fields}

\author{Vinay Kumar\aff{1}\corresp{\email{vinay.kumar@icts.res.in}}, and Pallavi Bhat\aff{1}}

\affiliation{\aff{1}International Centre for Theoretical Sciences, Tata Institute of Fundamental Research, Bangalore 560089, India}

\begin{document}

\maketitle

\begin{abstract}
 Magnetic reconnection, a fundamental plasma process, is pivotal in understanding energy conversion and particle acceleration in astrophysical systems. While extensively studied in two-dimensional (2D) configurations, the dynamics of reconnection in three-dimensional (3D) systems remain under-explored. In this work, we extend the classical tearing mode instability to 3D by introducing a modulation along the otherwise uniform direction in a 2D equilibrium, given by $g(y)$, mimicking a flux tube-like configuration. We perform linear stability analysis (both analytically and numerically) and direct numerical simulations to investigate the effects of three-dimensionality. Remarkably, we find that a tearing-like instability arises in 3D as well, even without the presence of guide fields. Further, our findings reveal that the 3D tearing instability exhibits reduced growth rates compared to 2D by a factor of $\int g(y)^{1/2} dy~/\int dy$, with the dispersion relation maintaining similar scaling characteristics. We show that the modulation introduces spatially varying resistive layer properties, which influence the reconnection dynamics. %
\end{abstract}

\section{Introduction}

Magnetic reconnection is a fundamental plasma process in the contexts of astrophysical, space and fusion plasmas \citep{zweibel_magnetic_2009}. This process occurs when oppositely directed magnetic field lines interact in the regions of high current density to break and reconnect, altering the magnetic topology and allowing for the rapid conversion of magnetic energy into kinetic and thermal energy. Magnetic reconnection can drive explosive phenomena such as solar flares, coronal mass ejections, and geomagnetic storms in Earth's magnetosphere \citep{shibata_solar_2011,ruan_fully_2020,burch_magnetic_2016}. It has played a pivotal role in regulating the dynamics of high energy  astrophysical environments, from the solar corona and interstellar medium to distant pulsar magnetospheres and black hole accretion disks \citep{zhang_generation_2023,fielding_plasmoid_2023,cerutti_modelling_2016,ripperda_magnetic_2020}. In particular, reconnection is  invoked to understand the particle acceleration and nonthermal emission in many systems \citep{cerutti_extreme_2012,sironi_relativistic_2014,guo_particle_2015,werner_extent_2016,brunetti_stochastic_2016,ghosh_magnetic_2024}.
Uncovering the mechanisms of reconnection is essential for explaining these high-energy events and for improving our understanding of plasma behavior across the universe.

Reconnection has been largely studied in two dimensions. It is thought to manifest in different ways : (i) spontaneously as tearing mode instability (ii) in steady-state known as the Sweet-Parker (SP) model and (iii) as turbulent reconnection. Tearing instability was first studied in the context of magnetic confinement in laboratory plasmas \citep{furth_finite-resistivity_1963,coppi_resistive_1976}. Since both tearing modes and the Sweet-Parker model lead to dimensionless reconnection rates that depend on a negative power of the Lundquist number ($S$), they could not explain observed time scales pertaining to solar flares. However, more recently, the discovery of the plasmoid instability has been considered to have solved the timescale problem \citep{loureiro_instability_2007,bhattacharjee_fast_2009,cassak_scaling_2009,uzdensky_fast_2010,pucci_reconnection_2013,comisso_general_2016}.
The plasmoid instability is fundamentally a re-emergence of the tearing instability in high-$S$ regimes and
arises asymptotically beyond $S\sim 10^4$, leading to bursty reconnection and the formation of small secondary magnetic islands called plasmoids \citep{samtaney_formation_2009, Landi_2015}.
Importantly, in its nonlinear regime, the plasmoid instability yields reconnection rates that are effectively independent of $S$ \citep{bhattacharjee_fast_2009}.

Complementary to this, the ideal tearing scenario proposed by \cite{pucci_reconnection_2013} predicts that thin current sheets can become unstable on ideal (Alfvénic) timescales when their aspect ratio scales as $S^{-1/3}$, leading to reconnection rates that are independent of resistivity. This prediction has been confirmed and extended through simulations and analysis in various geometries and conditions \citep{Landi_2015, Papini_2019, DelZanna_2016}.

Alternatively, there has been work that shows that 3D turbulence can also help in leading to fast reconnection \citep{lazarian_3d_2020}.
This turbulent model intrinsically uses the SP model for understanding the local reconnections and thus it is unclear if it needs some modification given that SP model is ruled out at values of $S$ higher than $\sim 10^4$.

In this work, rather than focusing on plasmoid instability or turbulent reconnection models, we aim to extend the study of the tearing mode instability into a fully three-dimensional context. This provides a complementary and necessary perspective for understanding reconnection in more realistic, inherently 3D systems. This has been approached in several ways, with two of the simplest being : (i) extending a 2D initial equilibrium into the third dimension and introducing 3D perturbations, (ii) incorporating a uniform guide field along the third dimension.

Approach (i) has been explored to demonstrate the occurrence of the kink instability, where the equilibrium current sheet buckles in response to 3D perturbations \citep{landi_three-dimensional_2008, oishi_self-generated_2015}. This buckling leads to nonlinear reconnection processes that can be faster than their two-dimensional counterparts. Reconnection setups are sensitive, even during the linear phase, to parameters such as whether the initial equilibrium is modeled using pressure balance or force-free fields, as well as the inclusion of a guide field \citep{landi_three-dimensional_2008}.
In the study by \citet{onofri_three-dimensional_2004}, which falls under category (ii), the presence of a guide field was found to stabilize the 3D instabilities observed by \citet{dahlburg_secondary_1992}, resulting in behavior that is closer to quasi-two-dimensional dynamics. They also observed that in the nonlinear regime, island coalescence leads to faster reconnection compared to the linear regime. Faster reconnection in the nonlinear phase appears to be a recurring finding across 3D tearing setups. For instance, \citet{wang_three-dimensional_2015} showed that introducing random perturbations, rather than specific mode perturbations, results in the formation of multiple tearing layers. These layers interact leading to faster reconnection.
Ultimately, many of these studies report that extending reconnection setups into three dimensions facilitates the generation of turbulence, which plays a crucial role in enhancing reconnection dynamics.

Another approach, taken primarily by solar physicists, is to investigate 3D field configurations that either have null points (where the magnetic field strength vanishes) or possess layers conducive to reconnection \citep{parnell_3d_2010}.
Studies of magnetic null points have identified distinct topological features, such as spine lines and fan surfaces \citep{priest_threedimensional_1995}.
These structures enable reconnection by directing magnetic flux along separatrix surfaces, which divide regions of differing magnetic connectivity \citep{wyper_non-linear_2014}.

However, many reconnection events, particularly in solar and astrophysical plasmas, occur in regions lacking null points \citep{demoulin_quasi-separatrix_1997}
In these cases, quasi-separatrix layers (QSLs) are thought to play a crucial role. QSLs are regions where magnetic field lines experience rapid connectivity changes, even without intersecting at null points. It is proposed that reconnection within QSLs can occur across a distributed region in the presence of intense current layers and, under certain conditions, exhibit bursty, explosive behavior similar to null-point reconnection 
\citep{aulanier_slip-running_2006,baker_magnetic_2009,kumar_magnetic_2021,mondal_reconnection-generated_2023}.

In this work, we explore the reconnection between anti-parallel, flux tube-like fields. This draws inspiration from the configuration used to examine vortex tube reconnection in \citet{melander_cut-and-connect_1989}, 
where the tubes, with cylindrical symmetry, are characterized by an axial component of the field with a radial dependence. We adopt a simplified version of this setup, which can be considered a modulation of the classic Harris sheet in the third dimension, described by $B_z(x)=\tanh(x)\sech^2(x)\sech^2(y)$.
Flux tube reconnection has been studied commonly with an intent to mainly explore the effect of twists and writhes in the field on the ensuing reconnection \citep{dahlburg_magnetic_1997,linton_reconnection_2001,wilmot-smith_magnetic_2007}. 
These configurations involving significant helicity are often referred to as flux ropes and exhibit complex 3D dynamics. In contrast, flux tubes without helicity are simpler, providing an idealized framework for examining fundamental aspects of magnetic reconnection. However, non helical flux-tube reconnection has received less attention in the literature compared to flux ropes \citep{linton_threedimensional_2003}.

In both cases—whether with or without helicity—previous studies have primarily considered flux tube interactions with either finite inclination angles or perpendicular orientations relative to each other. These interactions typically involve flattening of the tubes and formation of topologically complex structures during reconnection. In contrast, our study focuses on the simpler and less explored case of zero inclination angle, providing new insights into this idealized configuration. 

The further structure of the paper is as follows. In Section~\ref{lsa}, we describe the linear stability analysis using analytical and numerical approaches to understand the effects of modulation along the third dimension. Section~\ref{dns} outlines the setup of the direct numerical simulations used to test the theory and obtain the tearing mode growth rates, and Section~\ref{res} presents the results from the simulations comparing them with two-dimensional cases and discussing the impact of three-dimensional effects. Finally, Section~\ref{conc} summarizes our findings and suggests areas for future research.

\section{Linear Stability Analysis}
\label{lsa}

To investigate 3D effects in the tearing instability, 
we consider a 3D base state that consists of a modulation of the standard 2D equilibria, given by \(\mathbf{B}_0 = B_0 f(x)\hat{\mathbf{z}}\), in the third direction. 
The 3D initial configurations are obtained by modulating the corresponding 2D configurations along the third direction by a function $g(y)$, such that the 3D base state is then given by \(\mathbf{B}_0 = B_0 f(x)g(y)\hat{\mathbf{z}}\). To have a configuration of reversing magnetic fields along $x$, $f(x)$ is chosen to be an odd function. Moreover, as we shall show in Section \ref{dns}, a smoothly varying $g(y)$ with even parity such that $g(y)\rightarrow0$ as $\vert y \vert \rightarrow\infty$ produces a tube-like configuration of magnetic fields. We choose such a $g(y)$.

Since the magnetic field in such a configuration varies only across (and not along) itself, the associated magnetic tension, \(\mathbf{B}\cdot\nabla\mathbf{B} = 0\). The initial gas pressure distribution can then be chosen to cancel the magnetic pressure, thus ensuring that this initial configuration is in equilibrium. Here we have ignored the dissipation of the magnetic field, as is usually done in the tearing mode derivation under the assumption that the diffusive timescales are much larger than the tearing timescales.

We consider the full incompressible and inviscid MHD equations, 
\begin{align}
    \nabla\cdot\mathbf{u}&=0 \text{,}\label{eqn:solenoidality} \quad \nabla\cdot\mathbf{B} = 0 \text{,}
    \\
    \label{eqn:momentum}\dfrac{\partial}{\partial t} \nabla\times\mathbf{u} &=  - \mathbf{u\cdot\nabla \left(\nabla\times\mathbf{u}\right)} + \nabla\times\left[\left(\nabla\times \mathbf{B}   \right)\times \mathbf{B}  \right] \text{,}  %
    \\
    \label{eqn:induction}\dfrac{\partial \mathbf B}{\partial t} &= \nabla\times   \left( \mathbf{u}  \times \mathbf{B} \right) + \eta \nabla ^2\mathbf{B} \text{,}
\end{align} 
which comprise the solenoidality condition for the velocity and the magnetic field, and the evolution equations for the vorticity (from the momentum equation) and the magnetic field (the induction equation) respectively. Notice here that we are working in Alfv\'en units, that is, units wherein the plasma density, \(\rho\), is such that \(4\pi\rho = 1\).

The full MHD Eqs.~(\ref{eqn:solenoidality} - \ref{eqn:induction}) are linearized about the base states, \(B_0 f(x)g(y)\hat{\mathbf{z}}\) in terms of the perturbed magnetic fields, \( \mathbf{b}\) and the perturbed velocity, \( \mathbf{u}\). We work with vorticity instead of the velocity field to eliminate the pressure term from the momentum equation, making the analysis simpler ---  one does not need to track perturbations in the pressure. 
The linearized vorticity equation, component-wise, yields, 

\begin{align}
    \dfrac {\partial}{\partial t}\left(\partial_y u_z - \partial_z u_y\right) \notag & = B_0 \Big[f(x)g''(y)b_y + f'(x)g'(y)b_x+ f'(x)g(y)\partial_yb_x - f(x)g'(y)\partial_xb_x \\ & \notag -f(x)g(y)\partial_y\partial_xb_x-f(x)g(y)\partial_y\partial_yb_y-f(x)g(y) \partial_z^2b_y\Big],
    \\ 
    \dfrac {\partial}{\partial t}\left(\partial_z u_x - \partial_x u_z\right)\notag & = B_0 \Big[  - f'(x)g'(y)b_y - f(x)g'(y)\partial_xb_y - f''(x)g(y)b_x+ f'(x)g(y)\partial_yb_y \\ & \notag +f(x)g(y) \partial_z^2b_x +f(x)g(y)\partial_x\partial_xb_x+f(x)g(y)\partial_x\partial_yb_y\Big], \\
    \dfrac {\partial}{\partial t}\left(\partial_x u_y - \partial_y u_x\right) & = B_0 \Big[ \notag f'(x)g(y) \partial_zb_y+ f(x)g(y) \partial_z\partial_xb_y \\ &- f(x)g'(y)\partial_zb_x- f(x)g(y)\partial_z\partial_yb_x\notag\Big],
\end{align}

and the linearized induction equation, component-wise, is given by,
\begin{align}
    \dfrac{\partial b_x}{\partial t} & = B_0 f(x)g(y)\partial_z u_x 
    +  \eta    
    \left(
    \partial_x^2 b_x +
    \partial_y^2 b_x +
    \partial_z^2 b_x
    \right),\notag\\
    \dfrac{\partial b_y}{\partial t} & = B_0 f(x)g(y)\partial_z u_y +  \eta   
    \left(
    \partial_x^2 b_y +
    \partial_y^2 b_y + \partial_z^2 b_y
    \right),\notag\\
    \dfrac{\partial b_z}{\partial t} &= B_0 \Big[ f(x)g(y)\partial_z u_z-u_xf'(x)g(y)-u_yf(x)g'(y)\Big] 
    \notag\\ & + \eta  
    \left(
    \partial_x^2 b_z +
    \partial_y^2 b_z +
    \partial_z^2 b_z
    \right). \notag
\end{align}

We assume that any perturbed quantity is of the form
\[
    \psi = \psi(x,y) e^{ikz - i\omega t} \ ,
\]
 where \(\psi\) is a placeholder for either the perturbed magnetic field or the perturbed velocity, \(k\) is the wavenumber of the perturbation along $z$. 
 Using this ansatz, and the solenoidality conditions for the magnetic field and the velocity, we obtain

\begin{align}
    {-i\omega}\left( k^2 u_y-\partial_x\partial_y u_x - \partial_y^2 u_y\right) \label{eqn:linmomx}  & = ikB_0 \Big[ f(x)g''(y)b_y + f'(x)g'(y)b_x+  \\ & f'(x)g(y)\partial_yb_x  \notag   - f(x)g'(y)  \partial_xb_x -f(x)g(y)\partial_y\partial_xb_x \\ & -f(x)g(y)\partial_y^2b_y+f(x)g(y) k^2b_y\notag \Big],
    \\
    -i\omega \left(\partial_x^2 u_x \label{eqn:linmomy}+\partial_x\partial_y u_y-k^2 u_x \right)& =  ikB_0 \Big[ - f'(x)g'(y)b_y - f(x)g'(y)\partial_xb_y \\ & - f''(x)g(y)b_x  \notag   + f'(x)g(y)\partial_yb_y -f(x)g(y) k^2b_x  \\ & +f(x)g(y)\partial_x^2b_x+   f(x)g(y)\partial_x\partial_yb_y\Big]\notag, \\ -i\omega b_x & = ikB_0f(x)g(y) u_x + %
    \eta \left(\partial_x^2 b_x +\partial_y^2 b_x -k^2 b_x\right) \label{eqn:linindux},
    \\ -i\omega b_y & = ikB_0f(x)g(y) u_y + \eta %
    \left(\partial_x^2 b_y +\partial_y^2 b_y -k^2 b_y\right), \label{eqn:lininduy}
\end{align}

where \Eqs{eqn:linmomx}{eqn:linmomy} are obtained from the linearized vorticity equations and \Eqs{eqn:linindux}{eqn:lininduy} are obtained from the linearized induction equations respectively. Notice that \(u_z\) and \(b_z\) have been eliminated with the help of the solenoidality conditions. 

\subsection{Analytical approach to LSA}
\label{lin-analysis}

We adopt the approach in \citet{goldston_introduction_1995} for further analysis.
The main thing we focus on here is the derivation of the growth rate. The usual tearing 
mode analysis involves a boundary value problem. The current sheet is divided into three regions, 
the inner, the outer and the overlap regions. Resistive and inertial effects are negligible in the 
outer region and become important in the inner region where the resonant surface of $\mathbf{k}\cdot \mathbf{B}=0$ occurs and $f(x)\approx x$. All of this follows for the 3D case as well. 
The outer region equations can be used to completely characterize the instability parameter, 
\begin{equation}
\Delta^{'} = {\Big[ \frac{\partial \ln{b_x}}{\partial x} \Big]}^{0^+}_{0^+}\text{,} \quad \text{where }b_x=b_x(x,y)
\label{eqn:delprime}\text{.}
\end{equation}

We will take the $\Delta^{'}(x,y)$ as a given and proceed to examine the inner region equations which leads us to the growth rate of the 3D instability.  
From \Eq{eqn:linindux}, we can obtain an expression for $\partial_x^2 b_x$,  
\begin{equation}
\partial_x^2 b_x = \frac{-1}{\eta} \left( i\omega b_x + ik B_0 x g(y)u_x\right) \label{eqn:lin-ind},  
\end{equation}
where we have applied the standard considerations of $\partial_x^2 \gg k^2$ and $f(x)\approx x$ in the inner region.
Further, we have assumed $\partial_x^2 \gg \partial_y^2$ (See Appendix \ref{assump} for details) and thus dropped the corresponding term as well. 
Next we substitute the above into \Eq{eqn:linmomy}, to obtain,
\begin{align}
\frac{-\omega}{k} \left( \partial_x^2 u_x + \partial_x\partial_y u_y \right) =& B_0 \Big[ - x g(y) \left( \frac{i\omega b_x + ik B_0 x g(y)u_x}{\eta}\right) \notag \\
-& g'(y)b_y - xg'(y)\partial_xb_y + g(y)\partial_yb_y + xg(y)\partial_x\partial_yb_y \Big].
\label{eqn:lin-mom}
\end{align} 
We make an estimate of the characteristic length and velocity scales in the inner layer.
First we balance first term on the LHS with the second one from the RHS of the equation above, as one would have done in the 2D case as well. We can obtain the characteristic width of the inner resistive layer, 
\begin{equation}
x \sim \delta = \frac{\left(\eta \gamma \right)^{1/4}}{\left(B_0k g(y)\right)^{1/2}}.
\label{eqn:delta} 
\end{equation}
Here we have taken $\omega=i\gamma$ for growing modes. 

Next, we make the ansatz that $b_x \sim const. = \tilde{b}_x$ in the inner region. This ansatz is justified easily in the 2D case by showing that solutions of the kind $b_x \propto x^n$ (where $n \ge 1$) are excluded (see section 20.3 in \citet{goldston_introduction_1995}); turns out it can be extended to the 3D case as well.
Balancing the first two terms on RHS of \Eq{eqn:lin-mom}, leads to the velocity scale, 
\begin{equation}
\tilde{u}_x \sim \frac{i\gamma \tilde{b}_x}{kB_0 g(y) \delta}
\label{eqn:vel}   
\end{equation}
Now, we integrate \Eq{eqn:lin-ind} over the inner layer, 
we have, 
\begin{equation}
\Big[\partial_x b_x\Big]_{0-}^{0+} = \frac{1}{\eta} \int \left( \gamma \tilde{b}_x^2 - ik B_0 x g(y) u_x \right) dx 
\label{integ-b}  
\end{equation}

With \Eqs{eqn:delta}{eqn:vel}, we can transform the variables,  
$X \equiv x/\delta, V \equiv u_x/\tilde{u}_x$ and substitute into \Eq{integ-b} leading to, 
\begin{equation}
\frac{1}{\tilde{b}_x}\Big[\partial_x b_x\Big]_{x=0} = \frac{\gamma \delta(y)}{\eta} \int \left( 1+XV\right) dX.
\end{equation}
The integral on the RHS reduces to a function that remains dependent on $y$ which we will denote as $I(y)$. And the LHS term is basically the instability parameter $\Delta^{'}$ defined previously in \Eq{eqn:delprime}, which for simplicity we assume to be nearly homogenous along $y$.

We now substitute Eq 2.11 into Eq 2.14 to obtain, 
\begin{equation}
\Delta^{'} = \frac{\gamma }{\eta} \frac{(\gamma \eta)^{1/4}}{(kB_0 g(y))^{1/2}} I(y).
\end{equation}
where $I(y)=\int \left( 1+XV\right) dX$ is still a function of $y$ after the integration over $x$. Next we re-arrange the expression and identify that $\gamma$ needs to be independent of $y$ and thus we integrate over $y$ on both sides to obtain, 
\begin{equation}
\Delta^{'} \int g(y)^{1/2} dy = \frac{\gamma^{5/4} }{\eta^{3/4}(kB_0)^{1/2}} \int I(y) dy.
\end{equation}

Rearranging the equation, we can obtain the following dispersion relation, 
\begin{equation}
\gamma = \frac{{\Delta^{'}}^{4/5}\eta^{3/5} \left( k B_0\right)^{2/5} \int g(y)^{1/2}dy}{\int I(y)dy}. 
\label{finalgamma}
\end{equation}
The above is quite similar to the 2D dispersion relation obtained in the standard FKR regime pertaining to $\Delta^{'} \delta \ll 1$, except for the integrals over $y$. We will see in a  later section that indeed the 3D dispersion curves inferred from direct numerical simulations indeed do have the same behaviour as in 2D. The main difference arises from the integrals over $y$ and there is a reduction in the growth in this 3D case as compared to the 2D case by a factor of $\int g(y)^{1/2} dy/\int dy$. Here we have assumed that effect of modulation is negligible on the $I$ integral (this would be the case if the eigen function $b_x$ almost resembles its 2D counterpart  all along $y$; we show this in the next section).

Note that the above derivation hinges on the assumption that the modulation has a smooth or mild gradient along the third dimension. If this is not true then we cannot assume that $\Delta^{'}$ is homogenous along $y$ and effect of $I$ integral is negligible.

In particular, it is interesting to note that the three dimensionality of the initial equilibrium magnetic fields is such that the linear growth rate is modified from 2D to 3D by mainly a simple additional factor related to exactly the $y$ dependence (or the third direction dependence) in the initial field. This is attributed the simple nature (variable separable : $f(x)g(y)$) of the extension to 3D. The resulting growth rate expression can be understood in the following manner. 

If we were to extend the initial 2D field into 3D with no modulation in the third direction ($g(y)=1$), the 2D growth rates are recovered. However, due to the introduction of a modulation, the strength of the magnetic field becomes non-uniform which can affect the current sheet characteristics. In particular, we find that the characteristic length and timescales associated with the inner resistive layer are not uniform along $y$, whose effect precipitates a reduced growth rate. Again in a later section, we show that direct numerical simulations indeed confirm the growth rate reduction due to $\int g(y)^{1/2}dy$. 

\subsection{Numerical approach to LSA}\label{num-analysis}

The set of equations (\ref{eqn:linmomx}) - (\ref{eqn:lininduy}) can be written as a generalized eigenvalue problem which has the form \[
    \gamma \mathcal{M} v = \mathcal{L} v,
\]
with \(\mathcal{M}\) and  \(\mathcal{L}\) are linear operators acting on the eigenfunction \(v\), and \(\gamma\) is the corresponding eigenvalue. In our case, \(v\) comprises of \(u_x\), \(u_y\), \(b_x\) and \(b_y\), and we have substituted \(\gamma = -i\omega\) in the expectation of an unstable mode with \(\gamma>0\). 

We solve this generalized eigenvalue problem (EVP) numerically, using a spectral method with Fourier basis, to obtain the growth rate \(\gamma\) as a function of the wavenumber \(k\), and the corresponding eigenfunction. The full details of the solver can be found in Appendix \ref{evp_details}.

We choose a modified Harris sheet, as the unmodulated base state, with \(f(x) = 2.6 \tanh{(x)}\sech^2{(x)}\). The prefactor of $2.6$ ensures that the maximum value of the magnetic field strength is $1$, thus setting the Alfv\'enic velocity, \(v_A = 1\), making further normalizations simpler. Notice that the base state cannot be decomposed into a finite number of Fourier modes. This can lead to truncation errors, and naively proceeding with the numerical procedure described above produces oscillatory eigenfunctions, leading to a lack of convergent solutions. To overcome this, we apply an appropriate low-pass filter (details in Appendix \ref{lpf_details}) in Fourier space to the base state. This, by construction, gives us a base state suitable for Fourier decomposition.

\begin{figure}
    \centering
        \includegraphics[width = 0.8\columnwidth]{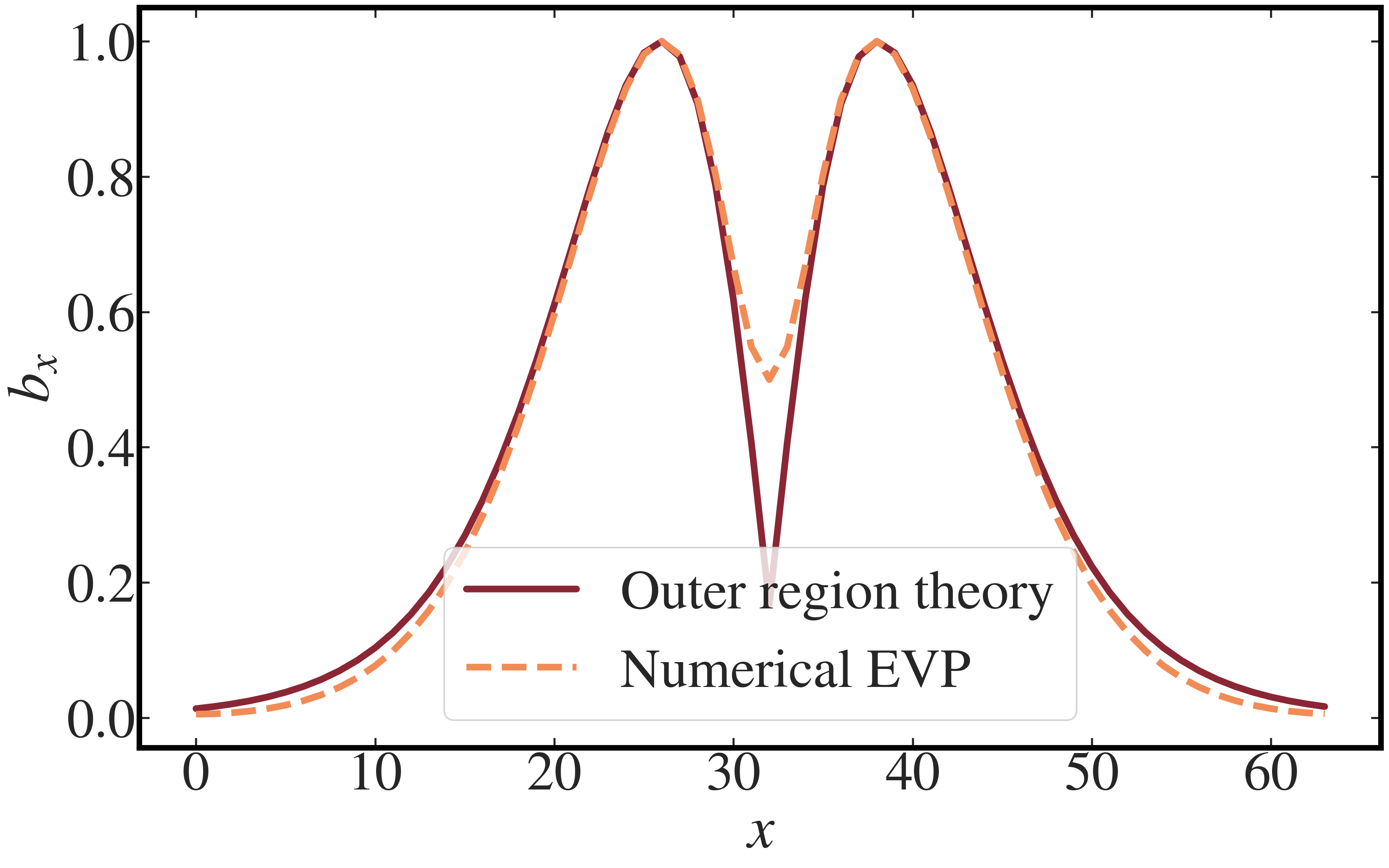} 
        
        \caption{ \label{fig:2D_efuncs_plot} Match between the numerically obtained eigenfunction for the 2D tearing instability
using the EVP solver and the expectation from outer region theory.} 
\end{figure}

For sanity check, we first obtained the eigenfunctions for the classical tearing instability in 2D. In this case, we have a 1D eigenvalue problem with \(v = [u_x,b_x]\). We confirm that our method reproduces the analytically calculated eigenfunctions and the dispersion relation. This is shown in \Fig{fig:2D_efuncs_plot}, which depicts the numerically derived eigenfunction along with the predictions of outer region theory.

        Further, we have verified that the Fourier filtering of the equilibrium profile does not alter the dispersion relation.  To establish this, we solved the EVP using a finite differencing algorithm with the unfiltered equilibrium. We did this only for the 2D cases since a finite difference algorithm requires a greater spatial resolution owing to poor convergence and running the EVP solver using the finite differencing algorithm becomes prohibitively expensive in 3D.

\begin{figure}
    \centering
    \includegraphics[width=0.75\linewidth]{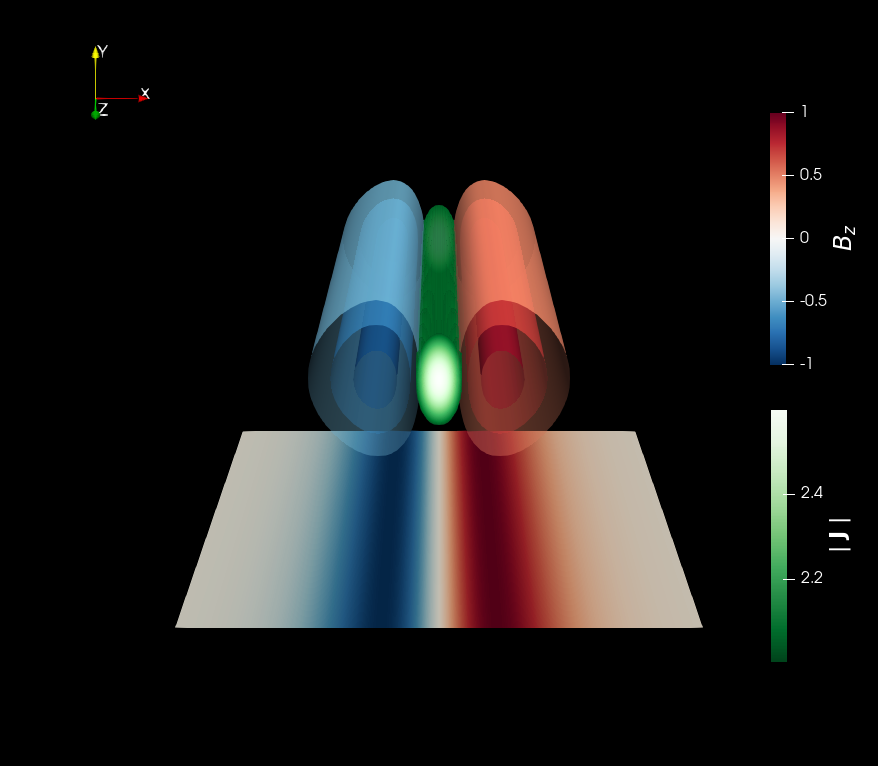}
    \caption{Equilibrium configuration of the magnetic field given by \Eq{eq:FullEq} with $\lambda = 1$. The red-blue colours show the $z$-component of the magnetic field and the region where the associated current density is greater than an arbitrary cutoff ($\vert\mathbf{J}\vert > 2$) is depicted in white-green colours. A slice of the unmodulated magnetic field is shown on the floor for comparison with the usual 2D tearing case. Notice that the modulation provides a tubular nature to the otherwise slab-like 2D configuration.}
    \label{fig:eq_isocont}
\end{figure}

We now present the results incorporating the modulation, \(g(y)\) into the equilibrium configuration. We choose \(g(y) = \sech^2{(y/\lambda)}\), thus modifying the equilibrium to, 
\begin{equation}
\mathbf{B} = f(x)g(y) = 2.6 \tanh{(x)}\sech^2{(x)}\sech^2{(y/\lambda)}  \hat{\mathbf{z}} \text{.} \label{eq:FullEq}    
\end{equation} 
The magnetic field and the current configuration arising from such a profile are shown in \Fig{fig:eq_isocont}. The parameter $\lambda$ controls the width of the modulation. 
An important length scale in the problem is the magnetic shear length \( a \), over which the field reverses, typically defined as \( B_z = f(x/a) \). Since we do not vary \( a \) in this work, we set \( a = 1 \) and use it to normalize all lengths throughout the analysis.
As we demonstrate in the next section, this setup closely mimics a system of reversing magnetic flux tubes with fields aligned along their axes.   

\begin{figure}
    \centering
    \includegraphics[width=0.9\columnwidth]{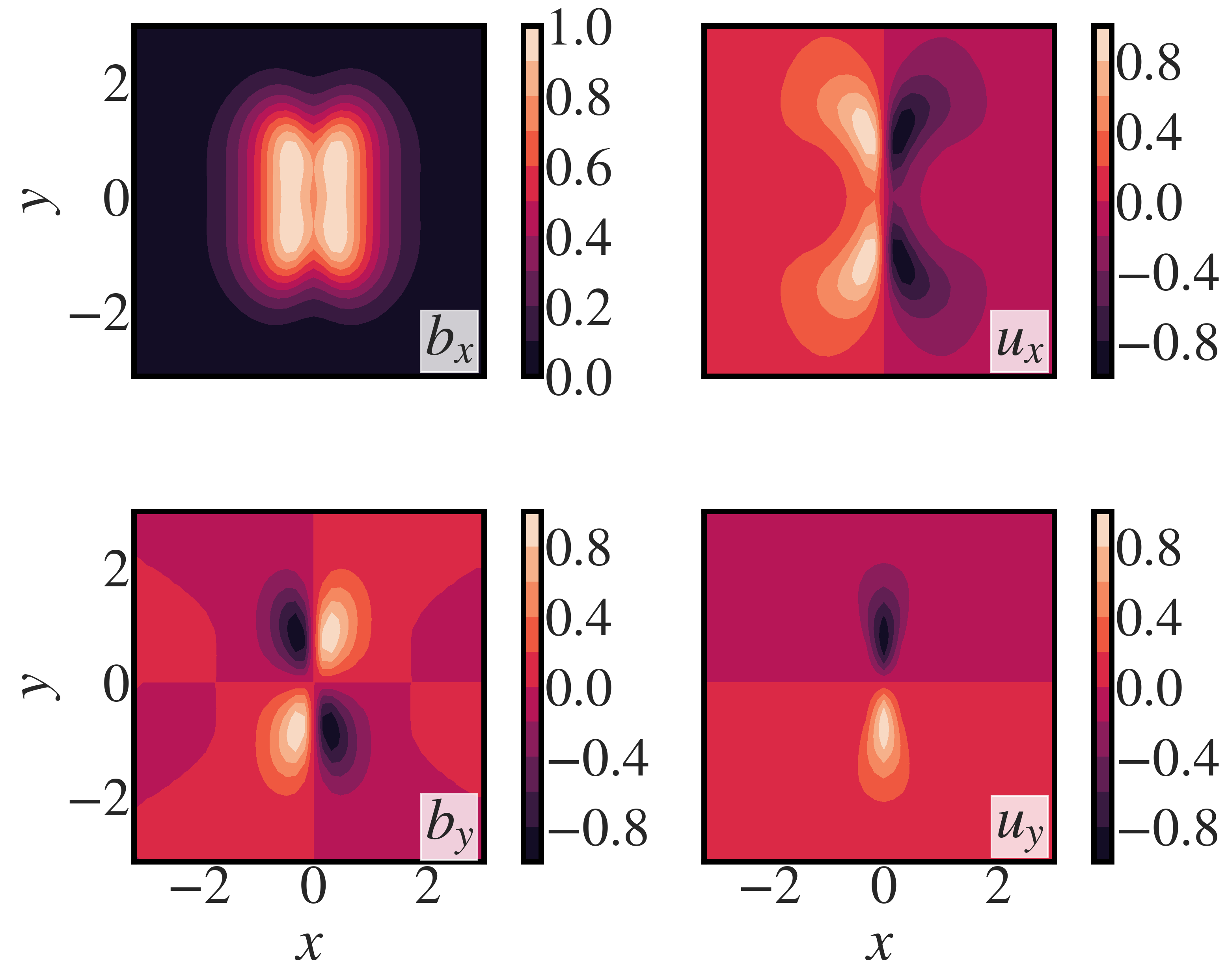}
    \caption{Eigenfunctions obtained from the numerical solution of the generalized eigenvalue problem with $\eta = 0.01$ and $k=0.7$. Since the eigenfunctions can only be calculated up to a scale, these were rescaled to have a spatial maximum of $1$. }
    \label{fig:EVP_efuncs}
\end{figure}

\Fig{fig:EVP_efuncs} illustrates the eigenfunctions computed for a particular set of parameters, $\eta=0.01$ and $k=0.7$. The spatial domain was taken as $(x, y) \in [-\pi, \pi) \times [-\pi, \pi)$ and we chose $N_x=N_y=64$.  The velocity eigenfunctions reveal flow fields converging at the origin, coinciding with the maximum equilibrium current density. The $x$-component of the perturbed magnetic field exhibits a double-humped profile, reminiscent of the purely 2D case, while the $y$-component of the magnetic field shows a quadrupolar structure. 

\begin{figure}
    \centering
    \includegraphics[width=0.75\linewidth]{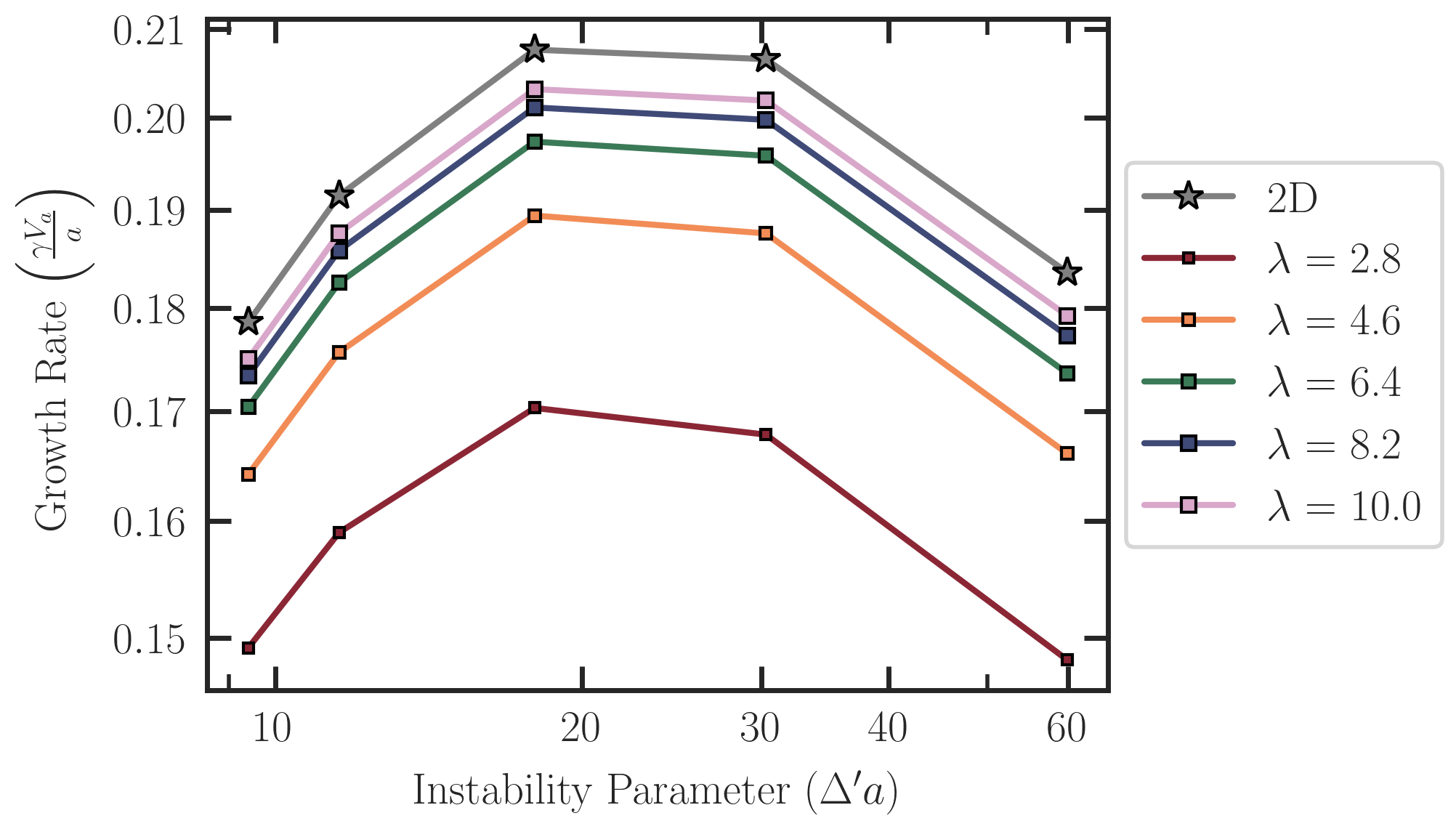}
    \caption{Dispersion curves with varying values of the modulation width $\lambda$.}
    \label{fig:disp_from_EVP}
\end{figure}

The full dispersion relation, as calculated using the eigenvalue solver is shown in \Fig{fig:disp_from_EVP}. Interestingly, the dispersion relations for the 3D cases (finite $\lambda$) bear close resemblance to the 2D dispersion curve, as expected from the theoretical considerations in Section~\ref{lin-analysis}. One must note here that the instability parameter \(\Delta ^ \prime\) in the 3D case is calculated using the same formula as in 2D. This is done in order to facilitate a comparison between the 2D and 3D cases.
The measured growth rates in the 3D cases, however, are smaller than their 2D counterparts.
Although the growth rate in 3D is smaller than that in 2D for any given wavenumber, the shape and the asymptotic scaling of the dispersion relation is similar in both cases. 

We validate these findings by performing direct numerical simulations in Section~\ref{dns}, and also highlight the impact of modulation on reconnection dynamics, including the scaling of growth rates and changes in the reconnection morphology.

\section{Direct Numerical Simulations}
\label{dns}

We perform simulations in both 2D and 3D to obtain the corresponding dispersion relations. The 2D simulations are necessary to validate the code by recovering the well-known tearing dispersion relation, and to have a benchmark for comparison with the 3D results. 

\subsection*{Numerical Setup}

The initial magnetic field configuration is described by \Eq{eq:FullEq}. Since this configuration is free of magnetic tension, we ensure equilibrium by choosing a suitable profile for the gas pressure, \(p_\text{gas}\), obtained by setting \(p_\text{gas} + p_\text{mag} = \text{const}\), where \(p_\text{mag} = B^2/2\) is the magnetic pressure. The fluid is perfectly stationary to begin with. 
The Lundquist number, $S$ can now be defined using a combination of the shear length, $a=1$, the Alfv\'enic velocity, $v_A = 1$, and the resistivity, $\eta$ as \(S=v_A a / \eta\). And the Alfv\'enic timescale is $\tau_A = a/v_A$.

To initiate the instability, perturbations of the form \(b_x = \sin(k^* z)\) are introduced in the \(x\) component of the magnetic field. For a given simulation, we introduce perturbations of only a single wavenumber, \(k^*\), in the \(z\) direction. This is done to estimate growth rates mode-by-mode, as is required to obtain the dispersion relation. Introducing perturbations with a multitude of wavenumbers would lead to the observation of only the fastest-growing mode, and one would not be able to get the dispersion relation.

We use a pseudo-spectral code, written using the Dedalus framework \citep{burns_dedalus_2020} to solve the usual visco-resistive MHD equations. The spectral expansion is done in Fourier basis which translates to periodic boundary conditions in all directions for both the 2D and the 3D case. We work with a resolution of \(N_x = 128\), \(N_y = 128\), and \(N_z = 64\), and the box size is \(L_x = 4\pi\), \(L_y = 4\pi\), and \(L_z = 2\pi /k^*\). This choice of \(L_z\) ensures that the perturbation of wavenumber \(k^*\) fits exactly in the box. As in Section~\ref{num-analysis}, a low pass filter is applied on the initial condition to ensure periodicity.  
 A 3/2 dealiasing scheme is implemented to avoid aliasing errors, and time stepping is done using a second-order Runge-Kutta scheme. Convergence tests were done to ensure that the results are not affected by the resolution. 

Because of finite, non-zero resistivity, the base state is not in equilibrium, rather, it diffuses out, as observed in \citet{landi_three-dimensional_2008}. This equilibrium diffusion (\( \eta \nabla^2 \mathbf{B}_0\)) is usually ignored in tearing mode analysis by assuming that the timescales of this diffusion are larger than the timescales of the tearing instability. This assumption holds well in the case of very small \( \eta \) as is typical of astrophysical systems. 
However, for our modest values of the Lundquist number, we find that the equilibrium diffusion timescale is comparable to the tearing instability timescale. This is especially true for the 3D case, where the equilibrium diffusion timescale is even smaller. 
This equilibrium diffusion leads to a slow decay of the base state, and so, the linear regime of growth is not sustained for long in our fully non-linear simulations.
To circumvent this, following \citet{landi_three-dimensional_2008} we add a constant term to the induction equation, \(-\eta \nabla^2 \mathbf{B}_0\), that rules out the equilibrium diffusion. This ensures that the base state does not decay and the tearing instability can be studied in the linear regime.

\section{Simulation Results}
\label{res}

\begin{figure}
    \centering
    \includegraphics[width=\columnwidth]{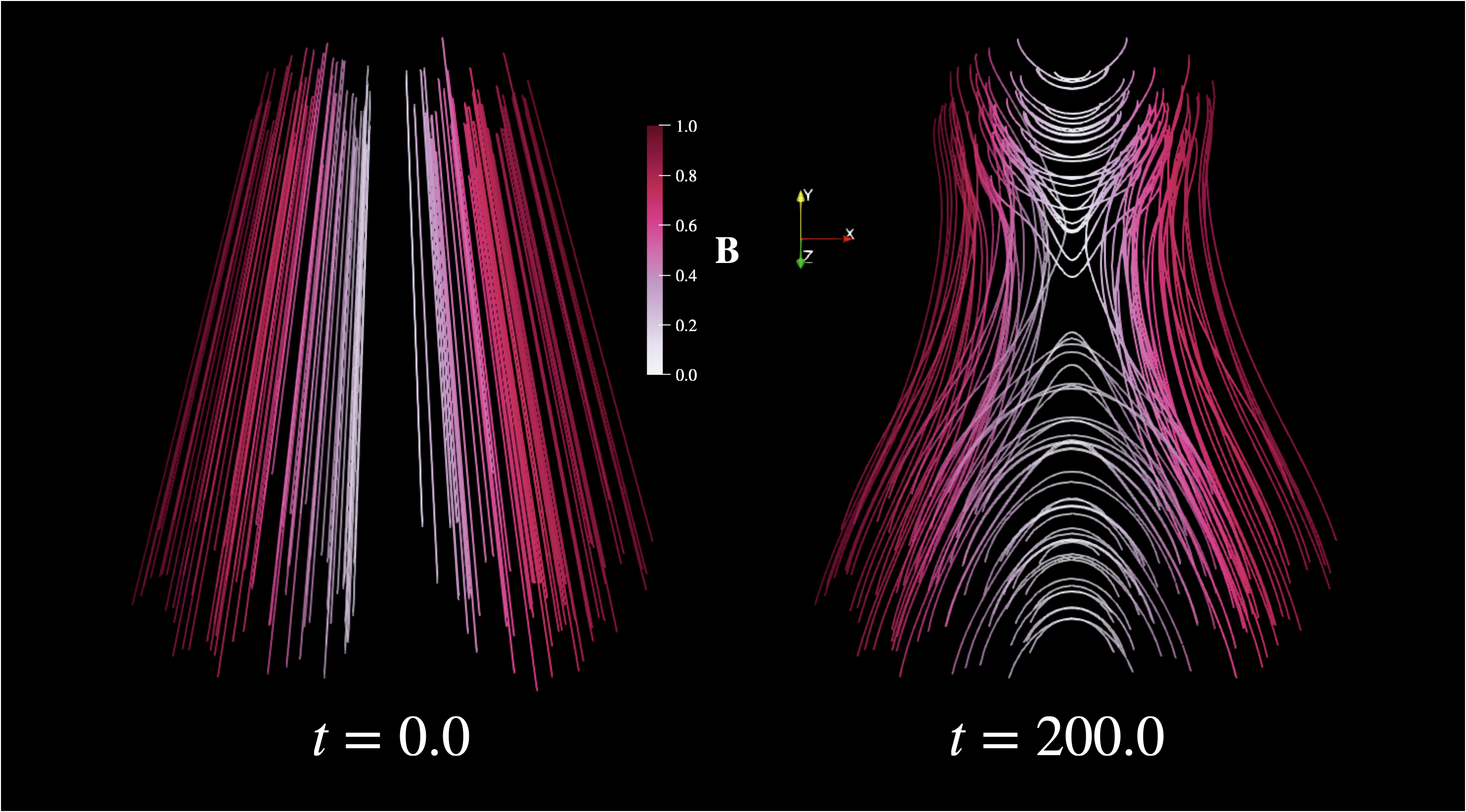}
    \caption{Streamlines of the magnetic field. The left figure shows the initial configuration and the flux-tube like structure is clearly visible. The structure at a later time is shown on the right. The colour represents the magnitude of the magnetic field.}
    \label{fig:pinkPlots}
\end{figure}

The evolution of the magnetic field driven by the 3D tearing instability is illustrated in \Fig{fig:pinkPlots}. The initially flux-tube-like magnetic fields converge, break, and reconnect to create new flux tubes. The behavior of these flux tubes closely resembles the dynamics of field lines observed in the 2D tearing instability.

\begin{figure}
    \centering
        \begin{minipage}{0.45\textwidth}
            \centering
            \includegraphics[width=\textwidth]{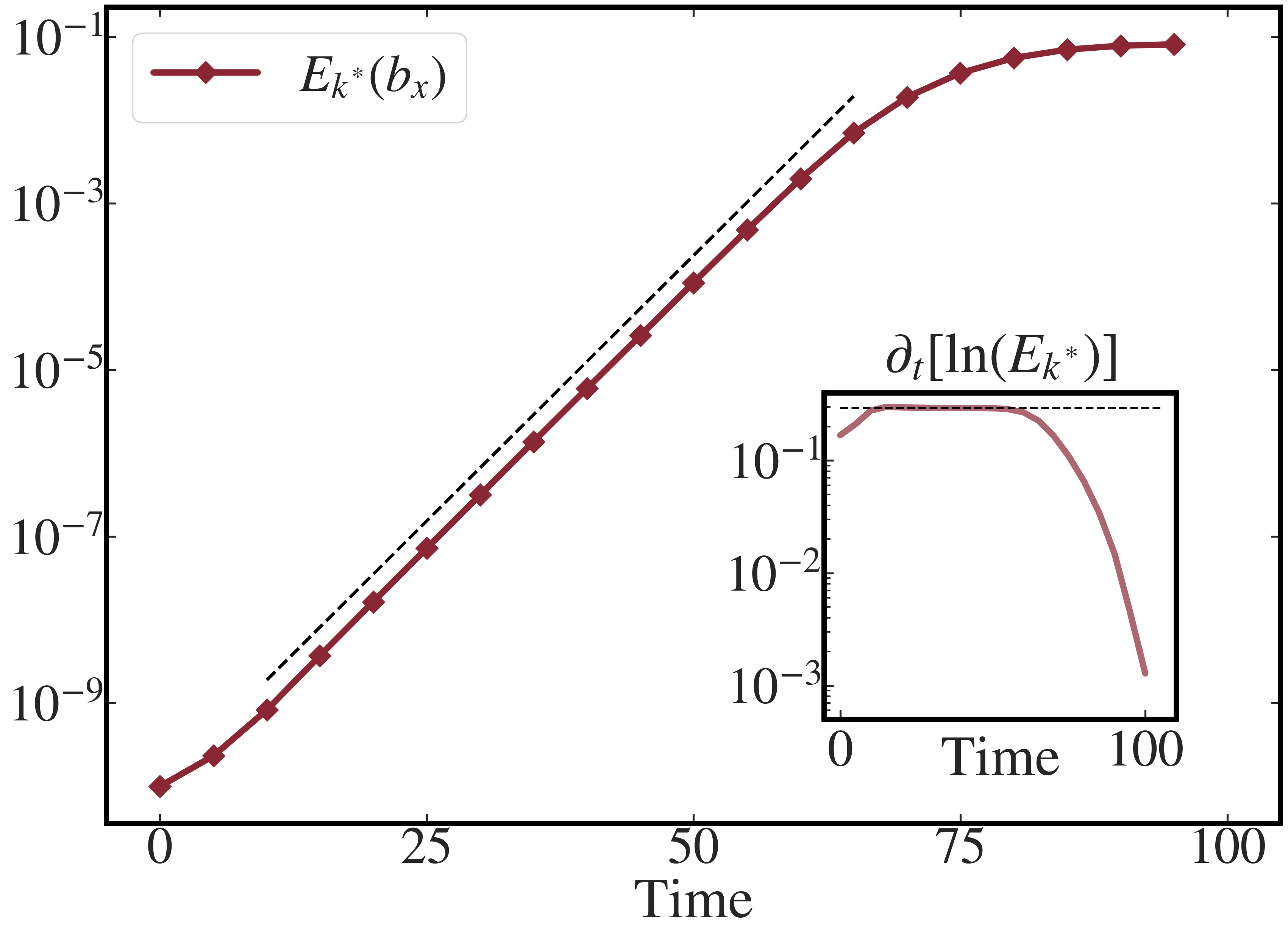}
        \end{minipage}
        \hfill
        \begin{minipage}{0.45\textwidth}
            \centering
            \includegraphics[width=\textwidth]{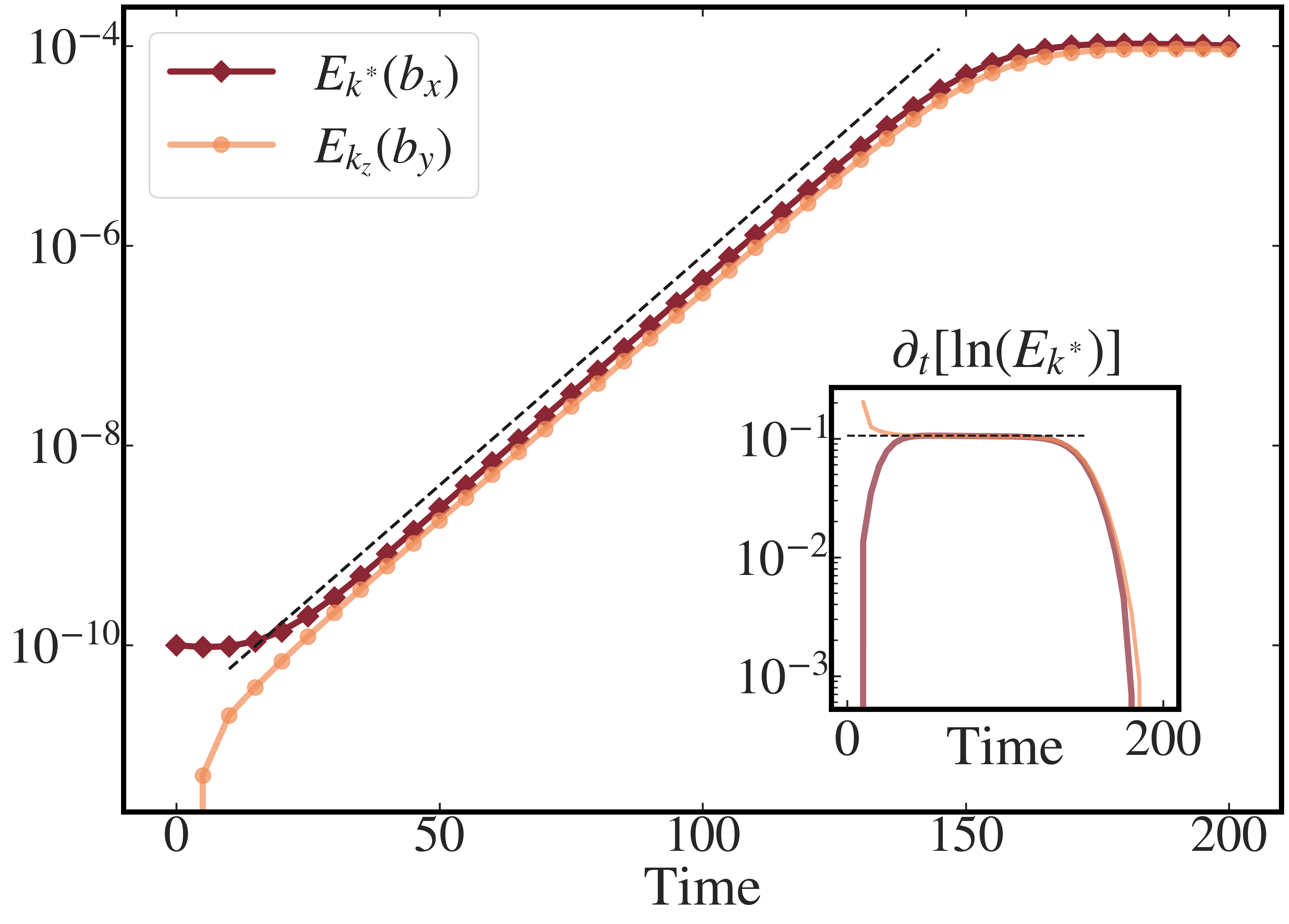}
        \end{minipage}
    \caption{Growth of perturbations in the 2D case (left) and the 3D case (right). The plot shows the linear growth of the energy in the unstable mode, \(E_{k^*}\), vs time (given in code units). The inset shows the evolution of the local slope, depicting a clean linear growth phase where the slope is a constant in time. The dashed line shows the fitted growth rate in both the plots.}
    
    \label{fig:2D_diag}
\end{figure}

The growth rate of the instability is measured by tracking the spectral energy in the \(x\) and the \(y\) components of the magnetic field for \(k = k^*\). This is given by

\[
E_{k^*}(b)=\mathlarger{\mathlarger{\sum}}_{k_{x},k_y}\left|\hat{b}\left(k_{x},k_y,k^*\right) \right|^{2} \text{,}
\]
where \(b\) represents the \(x\) or \(y\) component of the magnetic field, \(\hat{b}\) is the corresponding Fourier transform, and the summation is taken over all wavenumbers in the \(x\) and the \(y\) directions. The growth rate is then calculated by taking the local slope of \(\ln\left({E_{k^*}}\right)\) vs time. This shows a clean linear growth phase as shown in \Fig{fig:2D_diag}, where the local slope (shown in the insets) is a constant, as is expected in the linear regime of the tearing instability.

\begin{figure}
    \centering
    \includegraphics[width=0.95\linewidth]{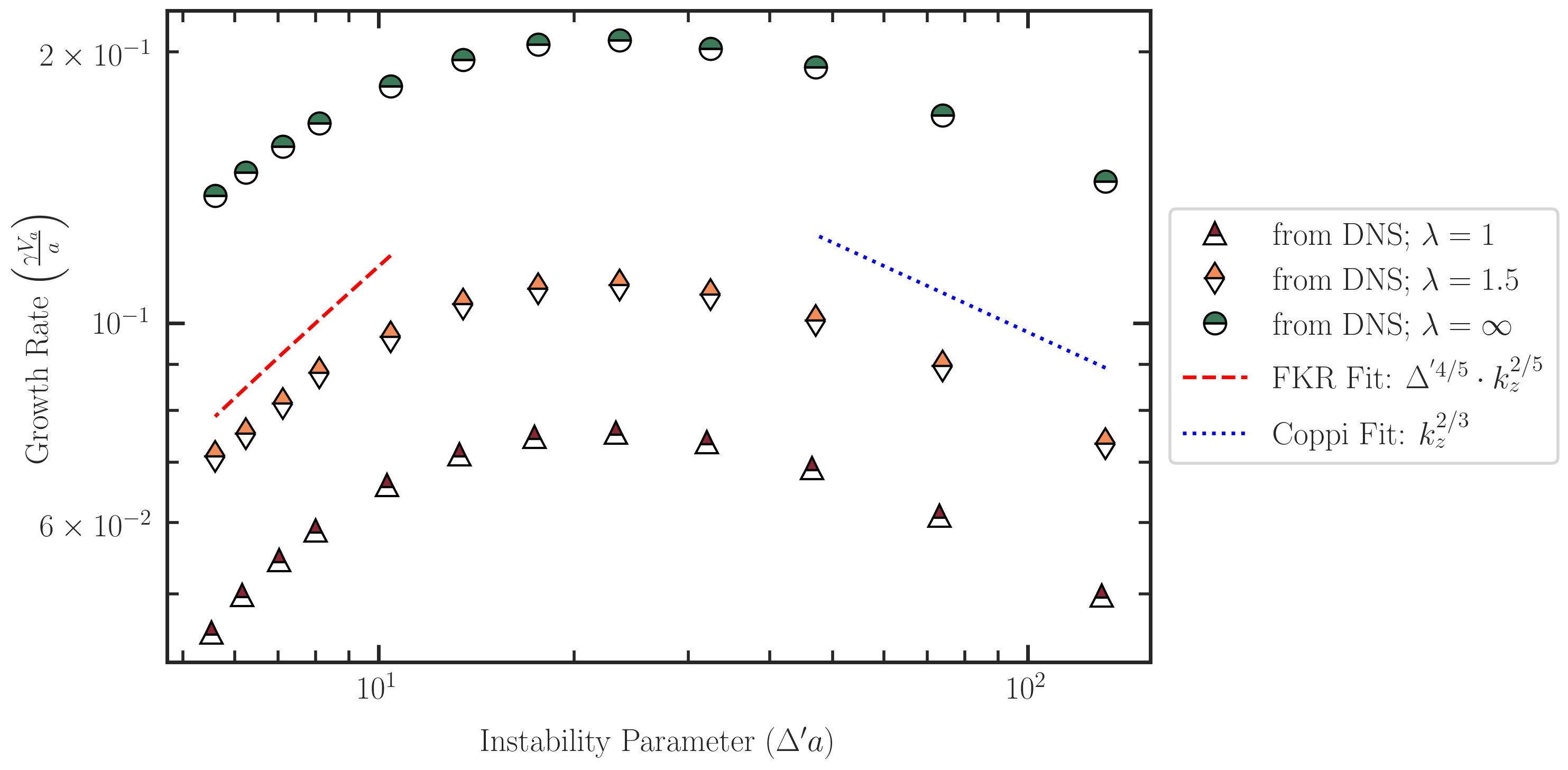}
    \caption{Dispersion relation in the 3D case with variation in the width of the modulation,  \(\lambda\). As before, the dashed lines show the asymptotic theoretical growth rates in the FKR and the Coppi regimes, and the points are the measured growth rates. The 2D dispersion relation with \(\lambda \rightarrow \infty\) is shown for reference.}

    \label{fig:3D_dispersion}
\end{figure}

The dispersion relation is obtained by measuring growth rates for different wavenumbers. These growth rates are plotted against the tearing instability parameter \(\Delta ^ \prime a\) in \Fig{fig:3D_dispersion}. The theoretical growth rate scalings in the FKR \(\Delta ^ \prime \delta \ll 1\) and the Coppi \(\Delta ^ \prime \delta \sim 1\) regimes are given in dot-dashed red and dotted blue lines respectively. The measured dispersion curves agree well with theoretical scalings. \Fig{fig:3D_dispersion} reaffirms our findings in the previous sections -- the growth rates in 3D are smaller but the shape of the dispersion curve and the fastest growing mode are independent of the width of the modulation. 

\begin{figure}
    \centering
    \includegraphics[width=0.75\columnwidth]{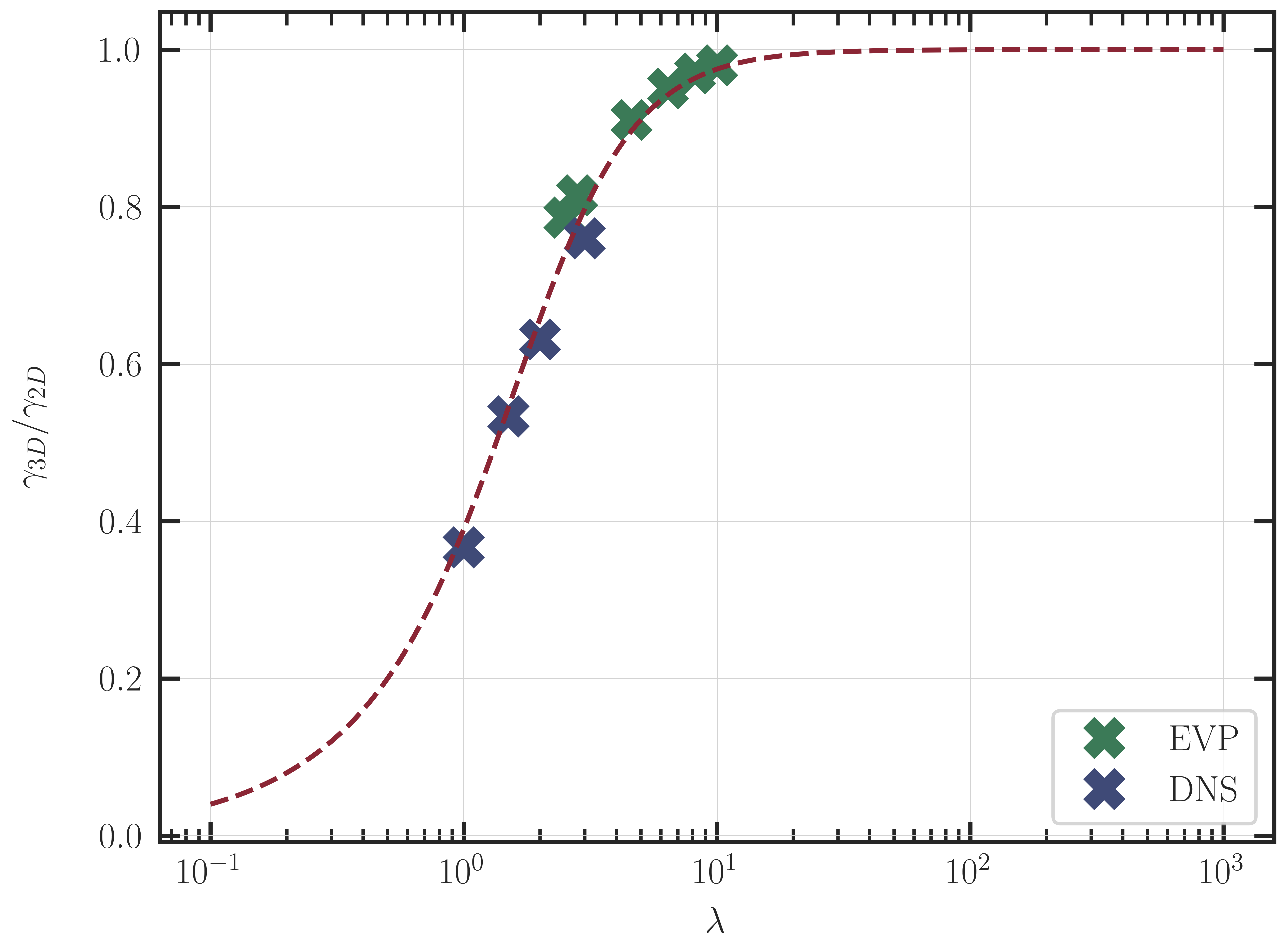}
    \caption{Effect of the modulation width \(\lambda\) on the growth rate. The plot on the top shows the ratio of the growth rate in 3D to that in 2D (or $\lambda=\infty$), \(\gamma_{3D}/\gamma_{2D} \) vs. the modulation width \(\lambda\), for a fixed wavenumber \(k^* = 0.7\). The blue crosses are measurements from the simulations and the green ones are from the eigenvalue solver. The maroon line is the theoretical prediction using \Eq{finalgamma}.}
        \label{fig:EffectOfMod}
\end{figure}

\begin{figure}
    \centering
    \includegraphics[width=0.75\linewidth]{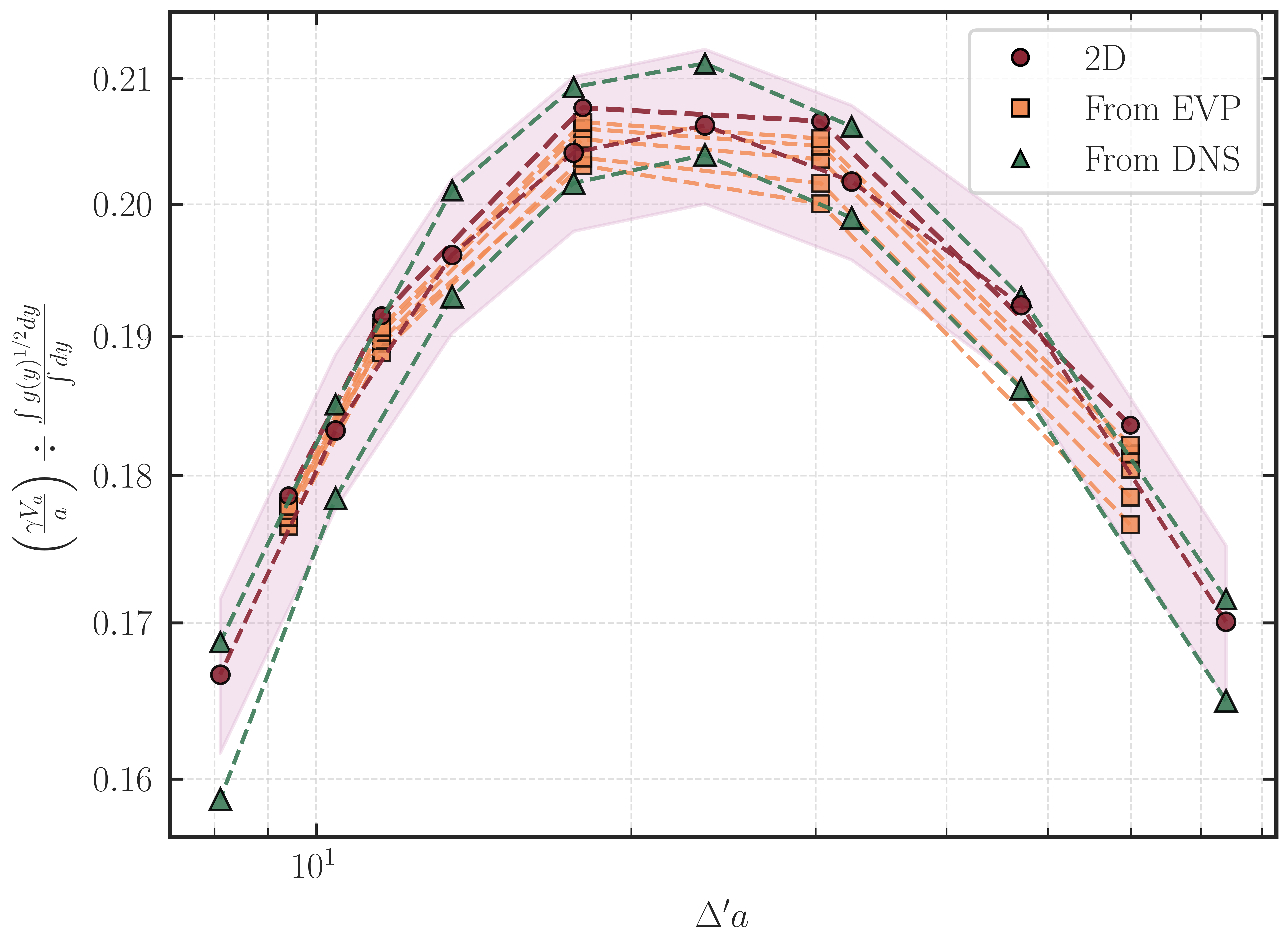}
    \caption{Collapse of the measured dispersion relation (from both the DNSs and the EVP) on top of the 2D dispersion curve when the corresponding growth rates are rescaled by the exact $\lambda$ dependent prefactor predicted in \Eq{finalgamma}.}
    \label{fig:collapse}
\end{figure}

As discussed in Section~\ref{lin-analysis}, the growth rates in the 3D case are affected by the modulation, as given in \Eq{finalgamma}. We now plot the effect of the modulation on the growth rate for a given wavenumber in \Fig{fig:EffectOfMod}. The maroon solid line shows the theoretical prediction of $\gamma$ for varying modulation width $\lambda$ in $g(y)$ using \Eq{finalgamma}. The 3D growth rates are reduced by a factor of $\int g(y)^{1/2}dy/\int dy$ from the 2D case of $\lambda=\infty,~g(y)=1$. We find that there is indeed a good match between the growth rates obtained from the eigenvalue solver, the simulations, and the theory. Further in \Fig{fig:collapse}, we show that the dispersion curves for different $\lambda$ fall on top of each other -- and the purely 2D dispersion curve -- when the measured growth rate is rescaled by the prefactor calculated in \Eq{finalgamma}. This collapse confirms that the primary effect of the modulation is to uniformly scale the growth rate across wavenumbers, without altering the shape of the dispersion relation. It validates our theoretical prediction that the spatial variation in $g(y)$ simply attenuates the overall instability strength.

\begin{figure}
    \centering
    \includegraphics[width=0.9\columnwidth]{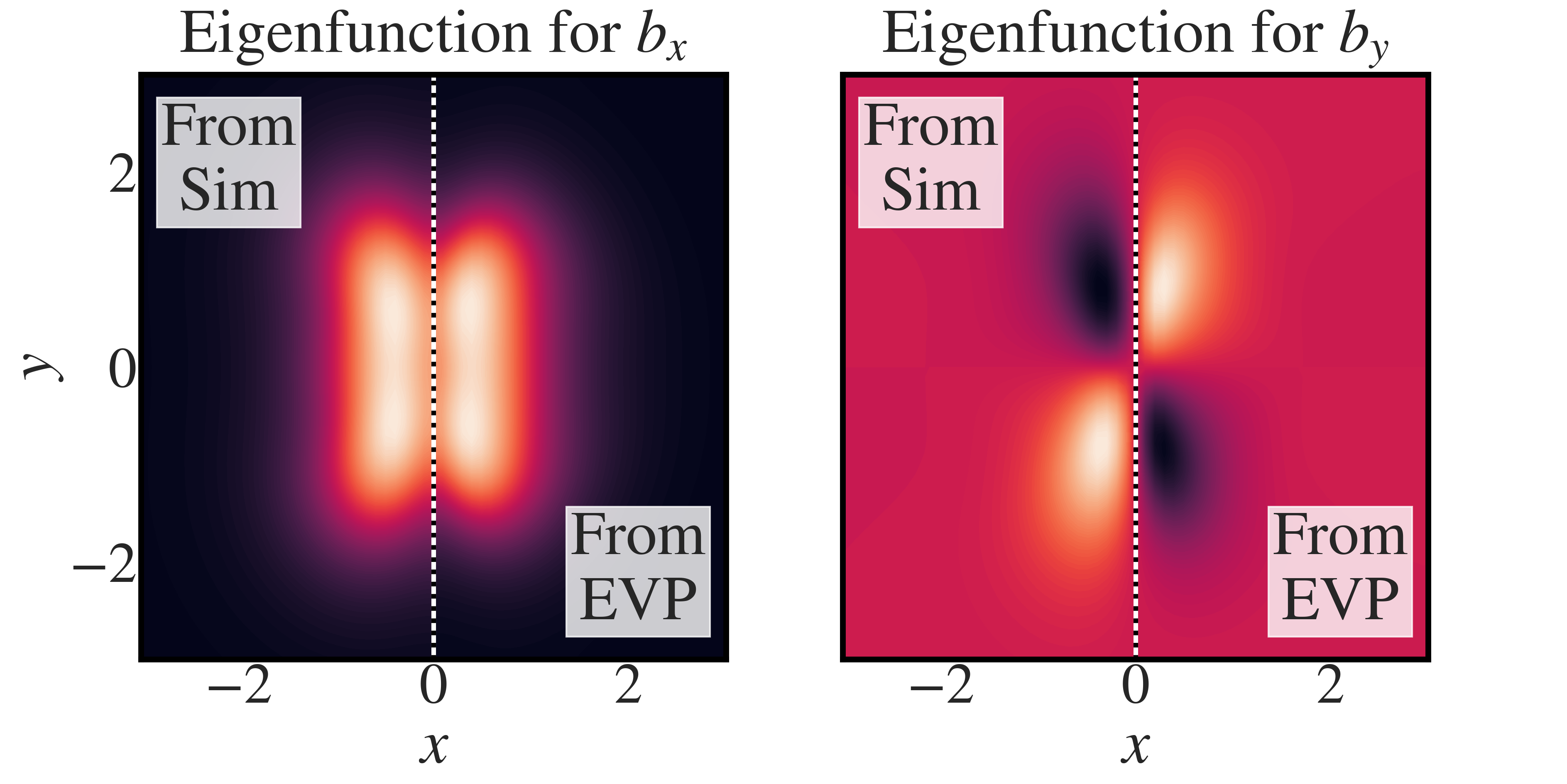}
    \caption{Comparison of the eigenfunctions obtained from the linear theory and the fully non-linear simulations. The eigenfunctions are shown for the same parameters, $\eta = 0.01$ and $k=0.7$. The left plot shows the $x$-component of the perturbed magnetic field, $b_x$, and the plot on the right shows the $y$-component of the perturbed magnetic field, $b_y$.}
    \label{fig:CompareEfuncsVertical}
\end{figure}

The magnetic field eigenfunctions can also be obtained from these fully non-linear simulations by taking a constant $z$ slice from the simulation domain. \Fig{fig:CompareEfuncsVertical} shows the so obtained eigenfunctions and they are in good agreement with those obtained from the linear theory in Section~\ref{num-analysis}. 

\begin{figure}
    \centering
    \includegraphics[width=0.55\columnwidth]{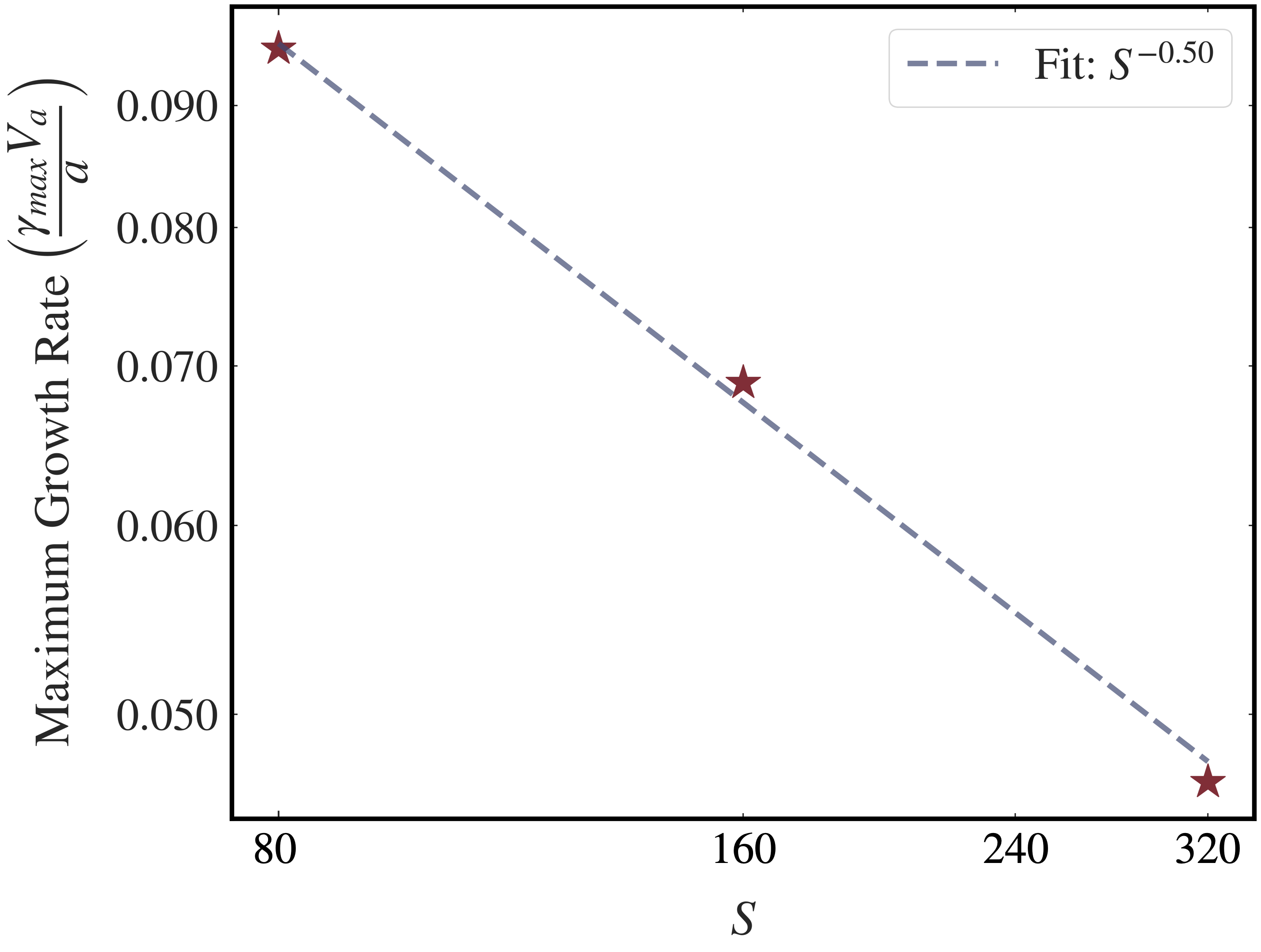}
    \caption{Scaling of the maximum growth rate with the Lundquist number \(S\). The maximum growth rate was obtained by  interpolating the individual dispersion relations. The solid line, which shows the \(S^{-1/2}\) scaling, is in good agreement with the data.}
    \label{fig:LundquistScaling}
\end{figure}

The Lundquist number scaling of the maximum growth rate in the 3D case is also similar to the 2D case. This was confirmed by performing 3 different suites of simulations with \(S = 80, \ 160, \  \text{and} \ 320\), and obtaining the full dispersion relation in each case. The maximum growth rate was estimated by interpolating the obtained dispersion relation on a finer \(\Delta^{'}\) grid and then taking the maxima. The scaling of the maximum growth rate with the Lundquist number is shown in figure \Fig{fig:LundquistScaling}.

\begin{figure}
    \centering
       
    \includegraphics[width=0.55\columnwidth]{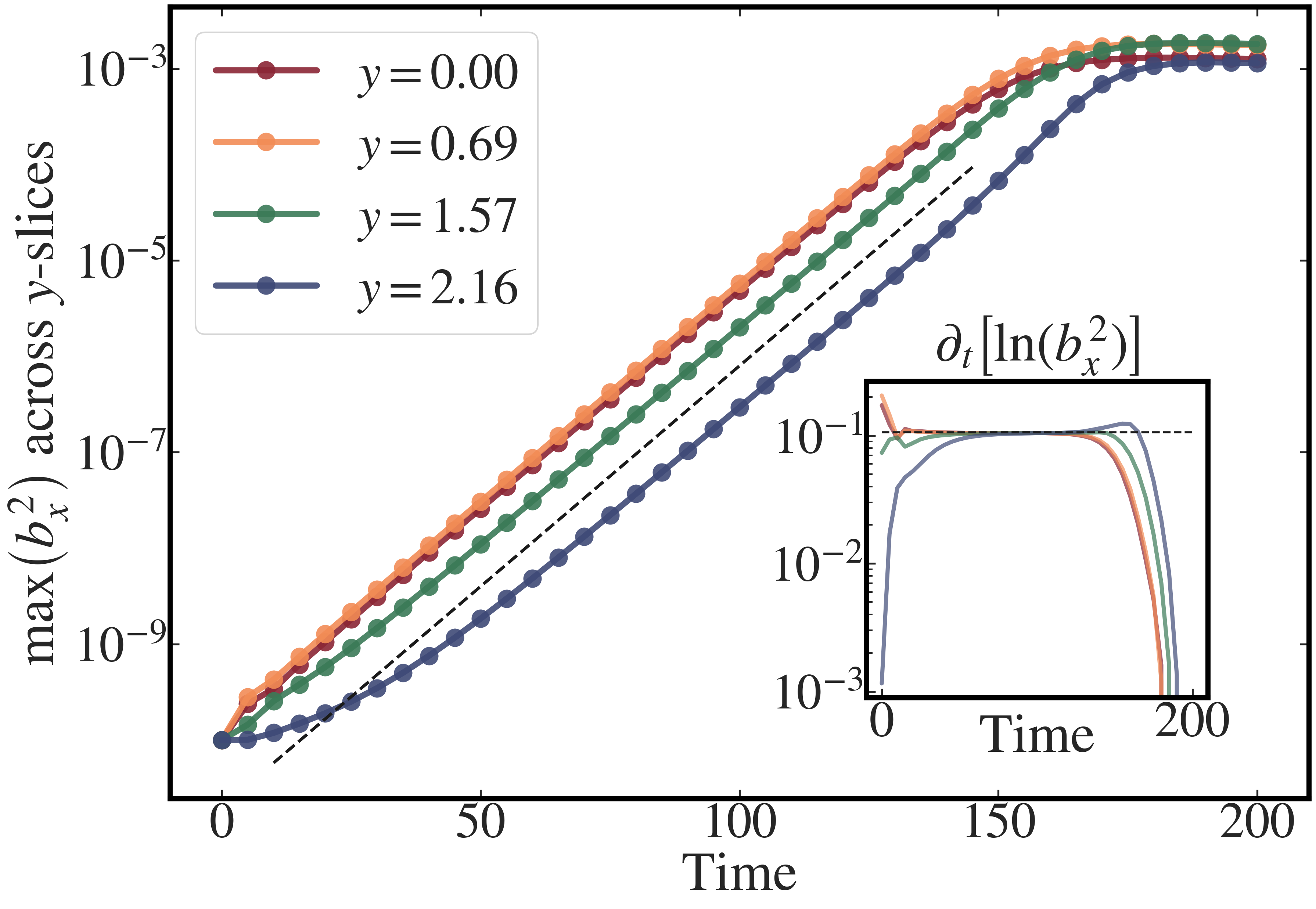}
        \caption{Growth of perturbations in different \(y\)-slices. The plot shows the maximum \(b_x^2\) across \(y\)-slices vs time. The growth rate is the same as that obtained from the spectral energy, indicating cross coupling between the slices.}
        \label{fig:acrossY}
        
\end{figure}

Given these similarities between the 2D and the 3D case, one must question if there is any impact of the third direction, or, is the 3D case simply a stack of different 2D tearing slices with different Alfvenic timescales, and hence different growth rates, which conspire to give a net smaller growth rate. To investigate this, we plot the growth of the perturbations in different \(y\)-slices. The maximum \(b_x^2\) across \(y\)-slices shows the same growth rate as shown in \Fig{fig:acrossY}, and this growth rate is also the same as that obtained from the spectral energy. This clearly signifies that all the \(y\)-slices have cross coupling, and hence, the 3D problem is not simply a stack of 2D slices.

\begin{figure}
    \centering
    \includegraphics[width=0.95\textwidth]{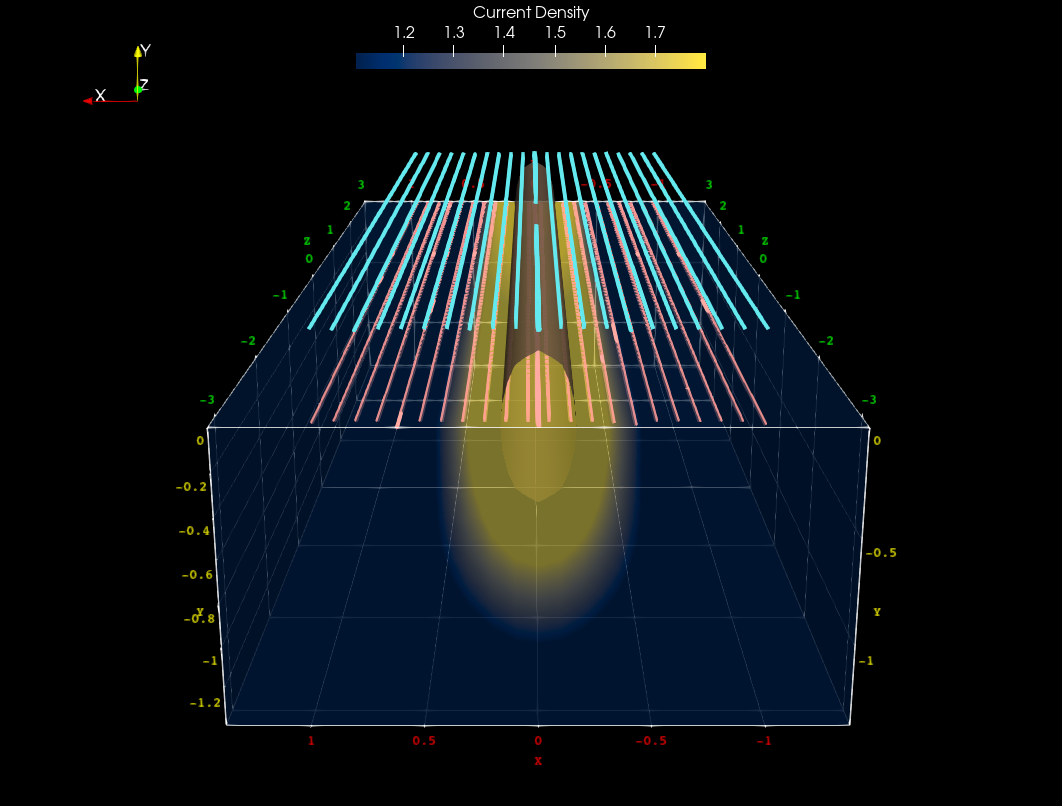} \\ 
    \includegraphics[width=0.95\textwidth]{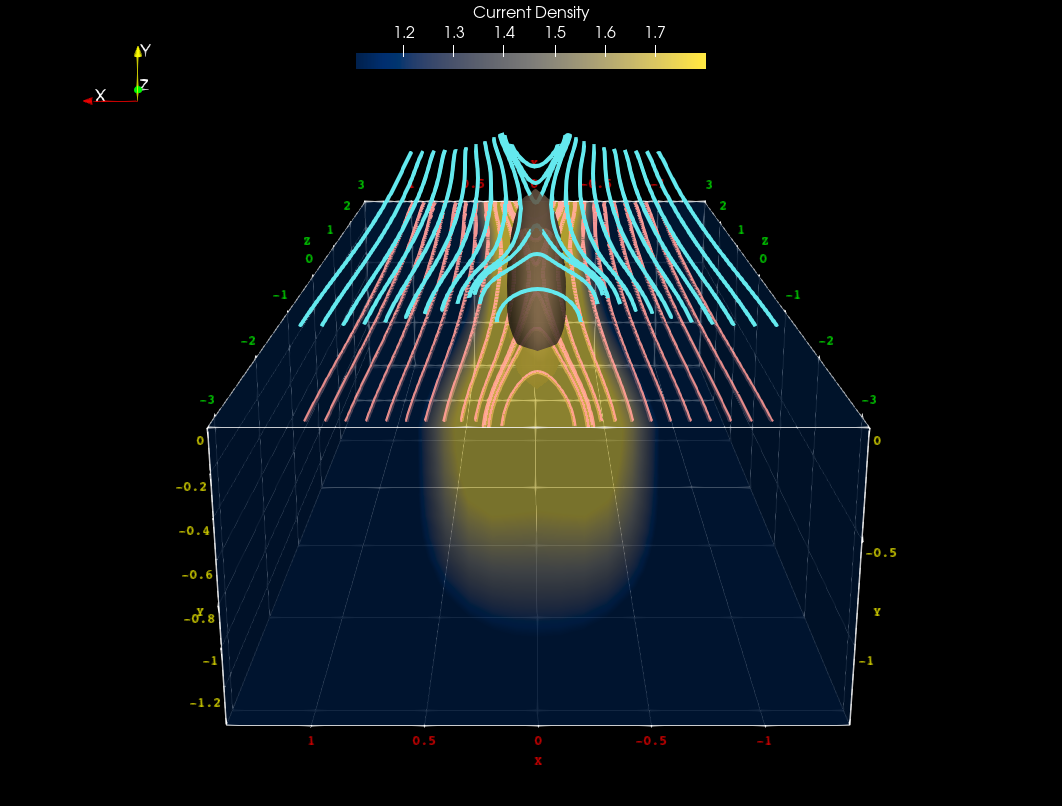}
    \caption{Magnetic field streamlines at two different times. Left: Initial state. Right: State at a later time in the linear growth stage. The volume rendering shows the current density and magnetic field lines are shown in pink (in the $y=0$ plane) and in cyan (off center plane, $y \neq 0$). The magnetic islands away from the $y=0$ plane appear bent in the $y$-direction.}
    \label{fig:current_streams}
\end{figure}

Next, we show in \Fig{fig:current_streams} various structures ensuing as a result of the  three-dimensional tearing instability. The volume rendering in the figure shows the current density, \(\mathbf{J}\) and an isocontour corresponding to a high value of \(\mathbf{J}\) is also shown in brown colour. 
The magnetic field lines in the $y=0$ plane are shown in pink, and bear close resemblance to the purely two-dimensional tearing instability. The cyan lines show the magnetic field in an off center plane. While these also have similarities with the 2D case, there is a clear three-dimensional structure to these lines, with a finite non-zero \(y\)-component of the magnetic field leading to magnetic islands that are bent in the $y$-direction.

In the purely two-dimensional scenario, a region of enhanced current density around an X-point with oval-shaped contours emerges along with the tearing mode and island formation. Analogously, in three dimensions, the current density contour reveals an oblong, ellipsoid-like structure as can be seen in the RHS plot in \Fig{fig:current_streams}. Taking the analogy further, in 2D tearing modes, the X-point draws magnetic field lines toward itself, facilitating reconnection at the X-point. This idea can also be extended to 3D. The pronounced current density (shown by the ellipsoid-like structure) is indicative of the magnetic null or the X-point close to the \(y=0\) plane responsible for pulling of the plasma containing field lines toward itself, causing the magnetic islands to be bent in the $y$-direction, and thereby giving rise to the observed 3D structure. Please see Appendix~\ref{nullp} for further details of island structures and their corresponding null points, along $y$.

\section{Conclusions and Discussions}
\label{conc}

In this paper, we studied a three dimensional  extension of the tearing instability. A straightforward 3D extension would consist of simply extending the standard 2D equilibrium into the third dimension and allowing for three dimensional perturbations. However, in such cases, the fastest-growing modes remain identical to their 2D counterparts, as per Squire’s theorem \citep{squire}. A more commonly explored 3D extension includes a uniform magnetic field along the third dimension, often referred to as a guide field. If the guide field is strong, then we recover the 2D behavior as seen in  reduced MHD. 

For our study, we considered a different 3D extension involving incorporation of a simple dependence of the field on the third direction, which we refer to as a modulation i.e. the 2D field, $B_z(x)$ was multiplied by a function, $g(y)$. In particular, we employed $B_z= \tanh(x)\sech^2(x) \sech^2(y/\lambda)$, which would represent a system with reversing magnetic flux tubes\footnote{This differs from flux rope configuration which consists of twisting around a central axis}. This can be considered as a simplified version of the vortex tube configuration in \citet{melander_cut-and-connect_1989}. 

Remarkably, we have found that this 3D equilibrium gives rise to tearing-like modes even without the presence of guide fields. Further, it turned out that this 3D configuration is amenable to tractable linear theory analysis. We derived a testable prediction: the impact of the modulation on the growth rate. Specifically, the linear growth rate is reduced by a factor of  $\int g(y)^{1/2}dy/ \int dy$. 
This reduction is attributed to the effect of the modulation on the properties of the inner resistive layer which are not uniform along the third dimension. 
Essentially, the reconnection retains a topologically 2D-like nature in such a system, but the three-dimensionality of the initial equilibrium—manifested through the modulation in the third direction—is reflected in the growth rate. It has been discussed in \citet{pontin_three-dimensional_2011}
 that reconnections are fundamentally different in 3D as field lines are not confined to a plane. However, we find that configurations as in our work allow for reconnections plane by plane with cutting and rejoining of field lines while maintaining a unified three dimensional character. 
In the ideal case, from the uncurled induction equation and Faraday's law, we have $\mathbf{E} + \mathbf{v}\times \mathbf{B}=\mathbf{\nabla}\Phi$. 
According to \citet{pontin_three-dimensional_2011}, it is possible to derive a smooth velocity field in 2D, as long as $\mathbf{E}\cdot\mathbf{B}=0$ and $(\mathbf{\nabla}\Phi)\cdot\mathbf{B}=0$, except at magnetic nulls. 
Thus, at X-points where the smoothness of the flow breaks down, we have reconnections. 
Further, they say that in 3D,  it is not possible to have $\mathbf{E}\cdot\mathbf{B}=0$ (due to the complicated geometry of the magnetic fields) and thus the conditions for flux freezing and magnetic topology conservation are more sophisticated. 
However, we find that our simple 3D reconnecting field configuration is such that $\mathbf{E}\cdot\mathbf{B}=0$ can still hold true ideally. Thus, we do observe a plane by plane X-point emergence allowing for 2D-like reconnections to occur.

In general, the evolution of a certain reversing magnetic field equilibrium depends sensitively on parameters such as whether it is in force-free or pressure based equilibrium and if there is a guide field or not \citep{landi_three-dimensional_2008}.
In particular, in our work, we have used pressure-balance based initial fields with no guide field. Importantly, we used mode-based perturbations (restricted to the $z$-direction) and thus did not study the most general fastest growing mode. 
While we haven't carried out a systematic study of this equilibirum with 3D random perturbations, we find that if the system is large enough along the third dimension, then the modulation doesn't inhibit the emergence of kink modes. 

We find that the scaling of the fastest growing mode with Lundquist number is similar to the 2D case of the tearing instability, $\gamma_{max} \sim S^{-1/2}$. This is noteworthy as the previous studies suggested that 2D-like reconnection persists only in the presence of a strong guide field. However, our results show that this $S^{-1/2}$ scaling in 3D can occur even without a guide field.

There are other questions that remain. How does the plasmoid instability manifest in such configurations? What happens if the 3D fields are helical? How would the reconnecting sheets be affected by turbulence? How does QSL reconnection fit into the picture? We plan to pursue some of these in the future.

\section*{Acknowledgements}

We acknowledge support of the Department of Atomic Energy, Government of India, under project no. RTI4001. The simulations were performed on the Contra cluster at the International Centre for Theoretical Sciences. We thank the anonymous reviewers for a thorough reading of the paper and their invaluable comments which have improved the presentation of our work. V. K. thanks Prayush Kumar, Saumav Kapoor and Vaishak Prasad for discussions with regards to the numerical eigenvalue solver. We also acknowledge illuminating discussions with the members of the ICTS AstroPlasma group.

\section*{Declaration of Interests}
The authors report no conflict of interests.

\section*{Author ORCID}
Vinay Kumar, https://orcid.org/0009-0008-2158-3774

\section{Appendix}

\subsection{Assumption of $\partial_x^2 \gg\partial_y^2$}
\label{assump}
\begin{figure}
          \centering
          \begin{minipage}{0.45\textwidth}
                \centering
                \includegraphics[width=\linewidth]{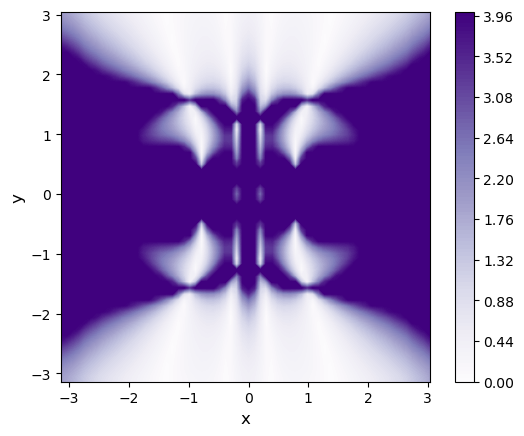}
          \end{minipage}
          \hfill
          \begin{minipage}{0.45\textwidth}
                \centering
                \includegraphics[width=\linewidth]{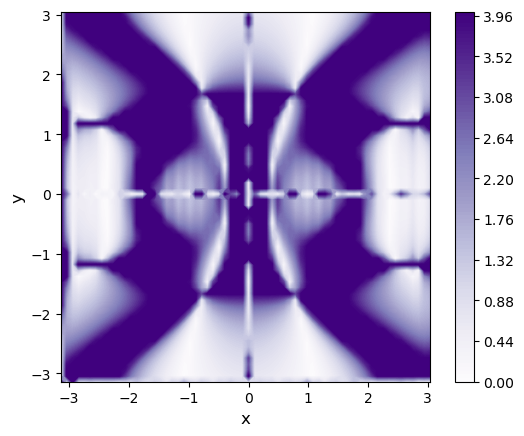}
          \end{minipage}
          \caption{Ratio of the absolute values of the double derivatives $\left( r = \dfrac{\partial_x^2}{\partial_y^2}\right)$ of $b_x$ (\textit{left plot}) and $b_y$ (\textit{right plot}), clipped at $r=4$ to avoid extreme dynamical range.}
          \label{fig:DD_b}
     \end{figure}

In Section \ref{lsa}, we had made a simplifying assumption that $\partial_x^2 \gg\partial_y^2$ in \Eq{eqn:linmomy}. We believe this is the case since we have considered an equilibrium that slowly varies in $y$. We posit that this behaviour should translate to the eigenfunction as well.
We show that this assumption indeed holds using the eigenfunctions obtained from solving the eigenvalue problem. Figure \ref{fig:DD_b} shows the ratio of the absolute values of the double derivatives $\left( r = {\partial_x^2}/{\partial_y^2}\right)$ of $b_x$ and $b_y$. As can be seen, the ratio is significantly greater than 1 in a large part of the domain, in particular, near the region of high magnetic shear, justifying our hypothesis.

\subsection{Details of the Eigenvalue Solver}
\label{evp_details}
The strategy is to express the linear operators, \(\mathcal{M}\) and  \(\mathcal{L}\), as matrices acting on a state vector, \(v\). Given these matrices, their eigenvalues, and the corresponding eigenvectors, can be found using publicly available schemes in many Python packages e.g. \texttt{SciPy \citep{2020SciPy-NMeth}, and PyTorch \citep{pyt}}.

We choose a numerical grid with $N_x$ and $N_y$ collocation points in the $x$ and the $y$ directions respectively. The eigenfunction now comprises of $N_x \times N_y$ values for each of \(u_x\), \(u_y\), \(b_x\) and \(b_y\). We flatten this eigenfunction to get the state vector, \(v\), with $4\times N_x \times N_y$ entries. The derivative operators are constructed by going to Fourier space, multiplying by appropriate powers of \(ik_x\) and \(ik_y\), and taking the inverse Fourier transform.
\begin{figure}
    \centering
        \includegraphics[width = \columnwidth]{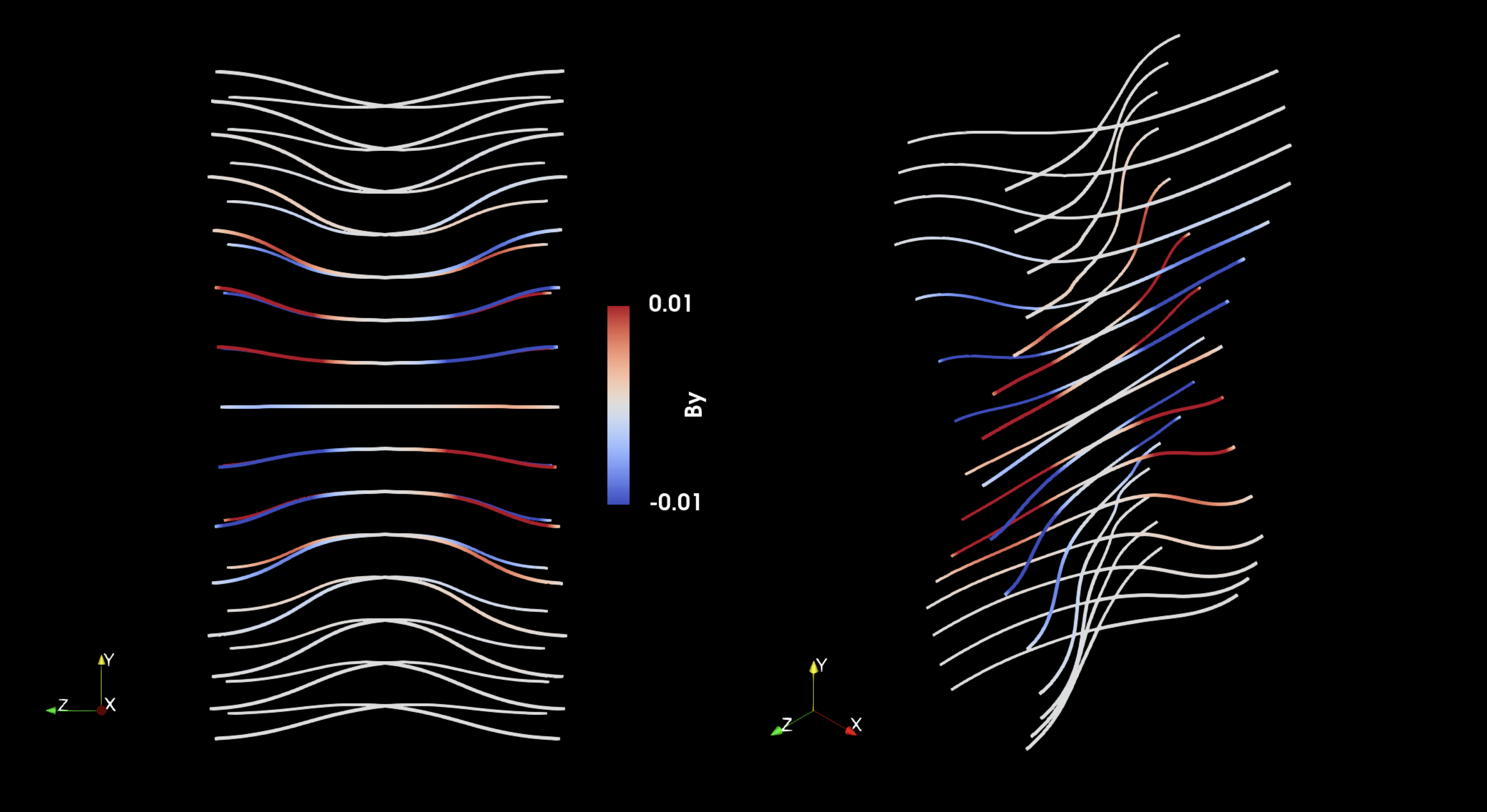} 
        
    \caption{Magnetic field lines in the full 3D case. The colours show the $y$-component of the magnetic field.}
    \label{fig:bendy}
\end{figure}

We then write a Python function with a dummy state vector as the argument which splits and reshapes it to give \(u_x\), \(u_y\), \(b_x\) and \(b_y\) over the 2D grid,  calculates the terms on the RHS of Eqs.~(\ref{eqn:linmomx}) - (\ref{eqn:lininduy}), and give these as output after flattening it back to a $4\times N_x \times N_y$ shape. The same is done for the LHS of Eqs.~(\ref{eqn:linmomx}) - (\ref{eqn:lininduy}), without the \(\gamma\) factor, which will be the eigenvalue obtained on solving the EVP. These functions take the state vector as the input and give as output a vector obtained by operating \(\mathcal{M}\) and  \(\mathcal{L}\) on the state vector. We now wish to obtain the matrix equivalent of these functions. This is done by first packaging the functions as \texttt{scipy.sparse.linalg.LinearOperator} objects, and then getting the matrix form of these objects by acting these on the identity matrix -- thus giving us the matrix form of \(\mathcal{M}\) and  \(\mathcal{L}\), each of size $4 N_x N_y \times 4 N_x N_y$.

The eigenvalues and eigenvectors for these matrices can now be calculated using standard methods. In anticipation of large runtimes due to the large size of the matrices, we resort to using a GPU-based Python library -- \texttt{pytorch}. We use the \texttt{torch.linalg.eig} package to get the eigenvalues and eigenvectors of the eigensystem comprised of the \(\mathcal{M}\) and  \(\mathcal{L}\) matrices. The numerical implementation of this full procedure has been made public in a GitHub repository --  \href{https://github.com/the21vk/SPEC-Tear.git}{SPEC-Tear}.

\subsection{Low-pass filtering and periodicity of the equilibrium}
\label{lpf_details}
We ensure periodicity of the equilibrium in the following way. The equilibrium is decomposed into its Fourier modes. Since the equilibrium is not strictly periodic, the Fourier coefficients for high wavenumbers are not zero, as would be the case for periodic functions. We now manually set the Fourier coefficients for wavenumbers above a chosen threshold to zero. The inverse Fourier transform of this gives us a perfectly periodic equilibrium. It is this process that we refer to as a low-pass filter -- since it gets rid of high wavenumber (frequency) modes, ensuring periodicity of the equilibrium.

\subsection{Magnetic island structure and null points}
\label{nullp}

The structure of the magnetic islands and the array of magnetic null points for the full 3D case is depicted by showing the magnetic field lines in \Fig{fig:bendy}. The colours indicate that $b_y$ is indeed non-zero, leading to a bent magnetic island structure. Note the symmetry of the structures across $y=0$ line. The most important aspect is that we recover 2D-like X-point sites which allow the classic ``cutting and rejoining" type of reconnections to occur plane by plane. More sophisticated geometries of magnetic fields may not allow for such simple manifestation of magnetic nulls.

\bibliography{3D_TMI}
\bibliographystyle{jpp}

    \makeatletter
    \def\fps@table{h}
    \def\fps@figure{h}
    \makeatother

\end{document}